\documentclass[
 reprint,
superscriptaddress,
 amsmath,amssymb,
 aps,
 prx,
]{revtex4-2}

\usepackage{subcaption}
\usepackage{bbold}
\usepackage{ragged2e} 
\DeclareCaptionJustification{justified}{\justifying}
\usepackage{graphicx}
\graphicspath{{figures/}}

\usepackage{dcolumn}
\usepackage{bm}
\usepackage{braket}

\usepackage[dvipsnames, table]{xcolor}

\usepackage{orcidlink}
\usepackage{hyperref}
\def\colorLinks{black}
\def\colorUrl{blue}
\def\colorCitations{blue}
\hypersetup{
  colorlinks=true,
  linkcolor=\colorLinks,
  urlcolor=\colorUrl,
  citecolor=\colorCitations
}

\usepackage{hyperxmp}

\usepackage{algorithm2e}
\RestyleAlgo{ruled}
\usepackage{algpseudocode}
\SetKwComment{Comment}{// }{}

\usepackage[normalem]{ulem}

\begin{document}

\preprint{APS/123-QED}

\affiliation{Dipartimento Interateneo di Fisica, Politecnico di Bari, 70126 Bari, Italy}                              
\affiliation{Dipartimento Interateneo di Fisica, Università degli Studi di Bari, 70126 Bari, Italy}                 
\affiliation{Istituto Nazionale di Fisica Nucleare, Sezione di Bari, Bari, Italy}                                   
\affiliation{Dipartimento di Fisica e Astronomia ``Galileo Galilei'', Università degli Studi di Padova, 35131, Padova, Italy}      
\affiliation{Istituto Nazionale di Fisica Nucleare, Sezione di Padova, Padova, Italy}

\title{\texorpdfstring{Percolation thresholds and connectivity in quantum networks}{}}

\author{Andrea De Girolamo\orcidlink{0009-0002-4529-0139}}\email{andrea.degirolamo@phd.unipd.it}
\affiliation{Dipartimento Interateneo di Fisica, Politecnico di Bari, 70126 Bari, Italy}
\affiliation{Dipartimento di Fisica e Astronomia ``Galileo Galilei'', Università degli Studi di Padova, 35131, Padova, Italy}
\affiliation{Istituto Nazionale di Fisica Nucleare, Sezione di Padova, Padova, Italy}

\author{Giuseppe Magnifico\orcidlink{0000-0002-7280-445X}}
\affiliation{Dipartimento Interateneo di Fisica, Università degli Studi di Bari, 70126 Bari, Italy}
\affiliation{Istituto Nazionale di Fisica Nucleare, Sezione di Bari, Bari, Italy}
\author{Cosmo Lupo\orcidlink{0000-0002-5227-4009}}
\affiliation{Dipartimento Interateneo di Fisica, Politecnico di Bari, 70126 Bari, Italy}
\affiliation{Dipartimento Interateneo di Fisica, Università degli Studi di Bari, 70126 Bari, Italy}
\affiliation{Istituto Nazionale di Fisica Nucleare, Sezione di Bari, Bari, Italy}

\date{\today}

\begin{abstract}
We study entanglement percolation in qubit-based planar quantum network models of arbitrary topology, where neighboring nodes are initially connected by pure states with quenched disorder in their entanglement. To address this, we develop a physics-informed heuristic algorithm designed to find a sequence of entanglement swapping and distillation operations to connect any pair of distant nodes. The algorithm combines locally optimal percolation strategies between nodes at a maximum distance of one swapping operation. If this fails to produce a maximally entangled state, it looks for alternative paths surrounding intermediate states within the process. We analytically find and numerically verify thresholds in quantum percolation, which depend on the initial network configuration and entanglement, and are associated with specific percolation strategies. We classify these strategies based on the connectivity, a quantity that relates the entanglement in the final state and the level of integrity of the network at the end of the process. We find distinct regimes of quantum percolation, which are clearly separated by the percolation thresholds of the employed strategies and vastly vary according to the network topology.

\end{abstract}

\maketitle

\section{Introduction}

The \textit{Quantum Internet}~\cite{Kimble2008QInternet,QIvision,Rohde_2021,QItech} is a visionary network infrastructure, enabled by entanglement~\cite{RevModPhys.81.865}, quantum repeaters~\cite{PhysRevA.59.169, PhysRevA.75.032310, Azuma_2023}, and quantum memories~\cite{Lvovsky2009qmemory}, which may allow long-distance on-demand quantum communications between users around the world. Potential use cases for quantum networks include information-theoretically secure communication~\cite{Pirandola:20}, distributed quantum sensing~\cite{Zhang_2021}, blind quantum computation~\cite{NielsenChuang2010,Cacciapuoti}, optical atomic clocks~\cite{Komar2014clock},
and very-long-baseline optical interferometry~\cite{PhysRevLett.123.070504,ZHuang}.

Due to the intrinsic fragility of entanglement, one aims to achieve long-distance quantum communications in a network topology by first generating entanglement between neighbor nodes, and then propagating it through quantum repeaters, with the support of quantum memories and error correction~\cite{PhysRevLett.78.3221, castro2025simulationentanglementbasedquantum}. For this purpose, it is crucial to select a robust topology~\cite{PhysRevA.97.012335,Mor_Ruiz_2025} and define entanglement routing protocols, in which entanglement is distributed between any pair of users connected through network nodes~\cite{routingGuha, Caleffi,SW2016,SW2019,SW2020,PhysRevLett.126.170501,Passarella,Battou,huang2024quantumentanglementpathselection,dawar2024quantuminternetresourceestimation,clayton2024efficientroutingquantumnetworks}. Through local operations and classical communications (LOCC) among nodes, maximal entanglement can be localized between distant nodes~\cite{Nielsen_1999,PhysRevLett.92.027901,Popp2005LocEnt,Chitambar_2014}. Even when a quantum network is initially composed of a large number of non-maximally entangled states, maximal entanglement can still be established between nodes that are separated by arbitrarily long distances under certain conditions. This phenomenon, known as \textit{entanglement percolation}~\cite{Acin2007EntPercolation,Perseguers2008PureQNets,PerseguersReport,Meng_2023,Meng2023PercolationReview}, enables long-range quantum correlations, which will be crucial for future quantum communication. 

In this work, we analyze entanglement percolation protocols that combine the elementary operations of entanglement swapping and entanglement distillation~\cite{Zukowski1993EntSwapping,PhysRevA.60.194, Bennett1996EntDistillation, PhysRevLett.83.1046, gu2025constantoverheadentanglementdistillation} to connect any given pair of distant nodes. In particular, we focus on regular 2D lattice models where neighboring nodes share pairs of entangled qubits in pure, non-maximally entangled states~\cite{sadhu2024practicallimitationsrobustnessscalability}. We analytically investigate percolation strategies and the conditions under which they are able to generate maximally entangled states between distant nodes. To study percolation in large quantum networks, we develop a physics-informed heuristic algorithm that identifies a sequence of entanglement swapping and distillation operations to connect any pair of distant nodes in a planar quantum network. Our results show that local quantum percolation strategies can be combined to generate a maximally entangled state between any pair of distant nodes. This holds if the average Schmidt value of the initial states -- directly related to entanglement -- remains below a certain threshold, dubbed the \textit{percolation threshold} (or, equivalently, Schmidt value threshold)~\cite{Malik2022threshold}. The price to pay to achieve maximal entanglement in quantum percolation is that a number of nodes will be temporarily disconnected from the network, as they help to transfer high-quality entanglement into other nodes. By measuring the entanglement in the final state and the level of integrity of the network at the end of the percolation process, we observe different percolation regimes which depend on the local percolation strategies employed at each step.

The paper develops as follows. In Sect.~\ref{sec:qnets_gen}, we introduce the basic entanglement manipulating operations, namely swapping and distillation. In Sect.~\ref{sec:analytics}, we provide an analytical description of the quantum percolation dynamics. In particular, we write a general expression for the Schmidt value resulting from any quantum percolation between nodes at a maximum distance of two, define the Schmidt value threshold associated with a local percolation strategy, and extend the framework for node pairs at higher distances via recursion. In Sect.~\ref{sec:pi_heuristics}, we describe the physics-informed heuristic algorithm for simulating quantum percolation over long distances. The algorithm first combines local percolation strategies, and then improves any resulting non-maximally entangled states by exploring alternative paths. In Sect.~\ref{sec:connectivity}, we define \textit{integrity} as a measure of how intact the network remains after percolation and \textit{connectivity} as an indicator of the quality of the percolation process, relating integrity to the entanglement of the final state. In Sect.~\ref{sec:results}, we present numerical results for two different quantum network topologies with fully connected unit cells. We analyze network connectivity, integrity, and final entanglement as functions of the initial Schmidt value distribution for systems with different levels of disorder. Finally, in Sect.~\ref{sec:conclusion}, we summarize our findings and discuss potential directions for future research.

\section{Entanglement manipulation in quantum networks}
\label{sec:qnets_gen}

Let us describe a quantum network model in which nodes represent stations, each with an arbitrary number of qubits, while links correspond to entangled pure states shared by qubits that belong to different stations. Without loss of generality, the quantum state of a pair of qubits can be expressed in its Schmidt form~\cite{Perseguers2008PureQNets}:
\begin{equation}
\label{eq:qstate}
    \ket{\lambda} = \sqrt{\lambda}\ket{00} + \sqrt{1-\lambda}\ket{11}
\end{equation}
with $\lambda \in \left[\frac{1}{2}, 1\right]$ and, consequently, $\lambda \geq 1 - \lambda$. As a measure of entanglement, we consider twice the smallest of the Schmidt coefficients:
\begin{equation}
\label{eq:ent}
E(\ket{\lambda}) = 2 (1-\lambda)
\end{equation}

Our objective is to solve the following task: for any pair of distant nodes $A$ and $B$ in a quantum network, find a path that connects $A$ and $B$ such that the entanglement of their final state is maximized. Generating entangled states between distant nodes in quantum networks requires modifying the structure of the network itself by means of entanglement-manipulating operations. A combination of such operations, namely entanglement swapping~\cite{Zukowski1993EntSwapping,PhysRevA.60.194} and distillation~\cite{Bennett1996EntDistillation, PhysRevLett.83.1046}, forms a quantum percolation strategy. In this section, we describe the dynamics of entanglement manipulating operations and assert the effect of both swapping and distillation on non-maximally entangled states.

\subsection{Entanglement swapping}

\begin{figure}[t]
    \centering
    \includegraphics[width=7cm]{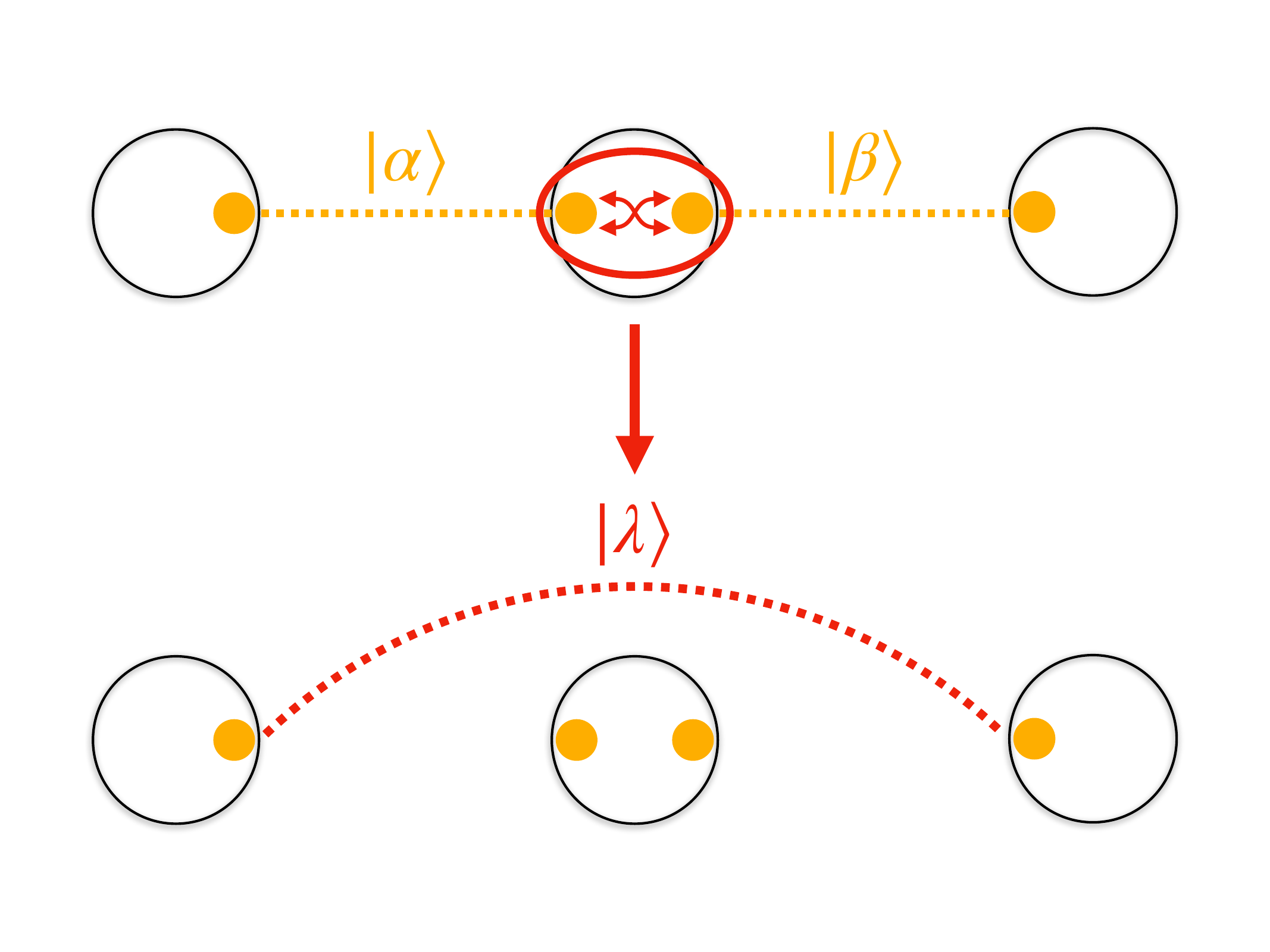}
    \caption{Graphical representation of an entanglement swapping operation, i.e., a combination in series of two entangled pairs.}
    \label{fig:swapping}  
\end{figure}

Consider two pairs of qubits, described by states $\ket{\alpha}$ and $\ket{\beta}$. If the states are in series, as in Fig.~\ref{fig:swapping}, \textit{entanglement swapping}~\cite{Zukowski1993EntSwapping,PhysRevA.60.194} maps the two states into one pure state by performing a Bell measurement on the middle qubits in Fig.~\ref{fig:swapping}, together with the support of LOCC. The entanglement of the post-measurement state $\ket{\lambda}$ from Fig.~\ref{fig:swapping} depends on the output of the measurement, but it has been proven~\cite{Perseguers2008PureQNets} that, in the case producing the minimally entangled outcome, a Bell measurement in the XZ basis yields the state
\begin{align}
\ket{\lambda_{SWAP}(\alpha, \beta)} = &\sqrt{\lambda_{SWAP}(\alpha, \beta)} \ket{00} + \nonumber \\ &\sqrt{1-\lambda_{SWAP}(\alpha, \beta)} \ket{11} \,
\end{align}
with
\begin{align}
\lambda_{SWAP}(\alpha, \beta) \geq 1 - \lambda_{SWAP}(\alpha, \beta)
\end{align}
and
\begin{align}
\lambda_{SWAP}\left(\alpha, \beta\right) 
=
\frac{1 + \sqrt{ 1 - 16 \alpha (1-\alpha) \beta (1-\beta) } }{2} 
\, .
\label{eq:lambda_swap}
\end{align}

Since this state yields the worst-case entanglement under the chosen measurement, we consider it a deterministic outcome for the purposes of our analysis. To study the entanglement of the post-measurement state when swapping non-maximally entangled initial states, we can express a quantum state, as per Eq.~\ref{eq:qstate}, in terms of the imperfection $\epsilon$ of its Schmidt value from the maximally entangled case:
\begin{equation}
    \ket{\lambda} = \sqrt{\frac{1}{2} + \epsilon}\ket{00} + \sqrt{\frac{1}{2} - \epsilon}\ket{11}
\end{equation}
with $\epsilon := \lambda - \frac{1}{2}$. 
By rewriting Eq.~\ref{eq:lambda_swap} as a function of $\epsilon_1 := \alpha -\frac{1}{2}$ and $\epsilon_2 := \beta - \frac{1}{2}$, we obtain:
\begin{align}
    \lambda_{SWAP}\left(\frac{1}{2} + \epsilon_1, \frac{1}{2} + \epsilon_2\right) = \frac{1}{2} + \underbrace{\sqrt{\epsilon_1^2 + \epsilon_2^2 - 4\epsilon_1^2\epsilon_2^2}}_{=:\epsilon_{SWAP}\left(\epsilon_1, \epsilon_2\right)}
    \label{eq:lambda_swap_eps}
\end{align}
When $\alpha = \beta = \frac{1}{2}$, entanglement swapping generates a maximally entangled state. However, when both states are not maximally entangled, that is, $\epsilon_1, \epsilon_2 > 0$, the post-measurement state yields less entanglement than either of the original states, as entanglement swapping amplifies the initial imperfections.

\subsection{Entanglement distillation}

\begin{figure}[t]
    \centering
    \includegraphics[width=7cm]{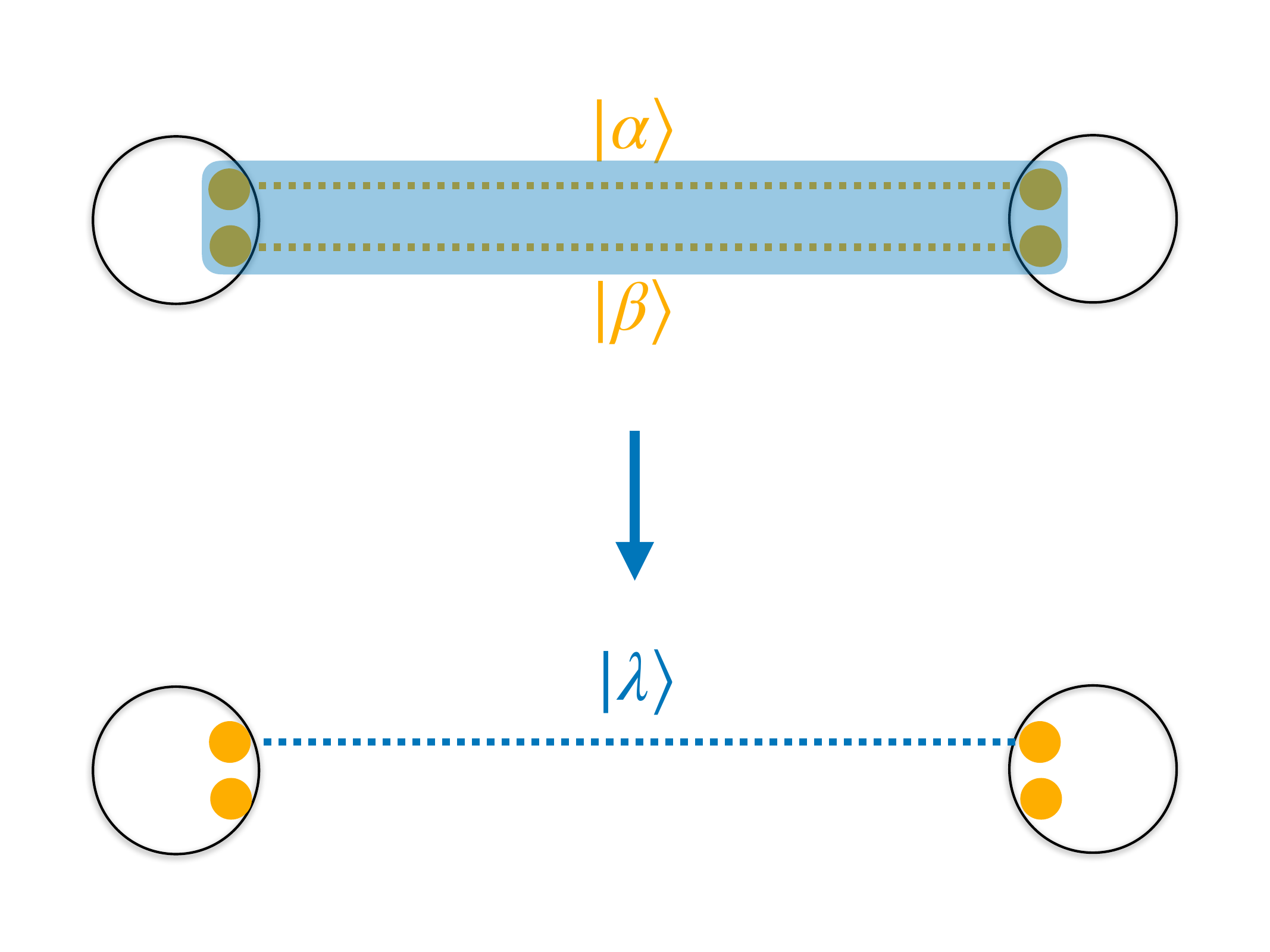}
    \caption{Graphical representation of an entanglement distillation operation, i.e., a combination in parallel of two entangled pairs.}
    \label{fig:distillation}
\end{figure}

Consider now the two entangled pairs arranged in parallel.
As shown in Fig.~\ref{fig:distillation}, it is possible to apply local measurements and operations, including classical communication, to create a single entangled state from a pair of states. This operation is called \textit{entanglement distillation}~\cite{Bennett1996EntDistillation}. The theory of majorization can be applied~\cite{Nielsen_1999, PhysRevLett.83.1046, NielsenVidal2001Majorization} to show that the most entangled state that can be obtained deterministically by distilling two pairs $\ket{\alpha}, \ket{\beta}$ is
\begin{align}
    \ket{\lambda_{DIST}(\alpha, \beta)} = &\sqrt{\lambda_{DIST}(\alpha, \beta)} \ket{00} + \nonumber \\ &\sqrt{1-\lambda_{DIST}(\alpha, \beta)} \ket{11} 
    \label{eq:state_dist}
\end{align}
with
\begin{align}
    \lambda_{DIST}(\alpha, \beta) \geq 1- \lambda_{DIST}(\alpha, \beta)
\end{align}
and 
\begin{align}
    \lambda_{DIST}(\alpha, \beta) = \max\left\{ \frac{1}{2} , \alpha \beta \right\}
    \label{eq:lambda_dist}
\end{align}

In Appendix~\ref{app:majorization}, we expand on majorization theory and prove the previous statement using Vidal's theorem~\cite{PhysRevLett.83.1046}. Rewriting $\lambda_{DIST}$ as a function of $\epsilon_1$ and $\epsilon_2$, as done previously for entanglement swapping, yields:
\begin{align}
    \lambda_{DIST}&\left(\frac{1}{2} + \epsilon_1, \frac{1}{2} + \epsilon_2\right) = \nonumber \\
    &\frac{1}{2} + \underbrace{\max\left\{0, \epsilon_1\epsilon_2 + \frac{1}{2}\epsilon_1 + \frac{1}{2}\epsilon_2 - \frac{1}{4}\right\}}_{=:\epsilon_{DIST}\left(\epsilon_1, \epsilon_2\right)}
\end{align}

Unlike in the case of entanglement swapping, the state $\ket{\lambda_{DIST}(\alpha, \beta)}$ resulting from distillation improves upon the entanglement of its parent states. In fact, distillation can produce maximally entangled states starting from two non-maximally entangled ones, up to some imperfection in the entanglement of the initial states.

\section{Analytical description of quantum percolation dynamics}
\label{sec:analytics}

An efficient entanglement percolation for a quantum network is a combination of entanglement swapping and distillation operations performed on the initial states of the network, with the aim of creating a perfectly entangled state between two nodes $A$ and $B$. In principle, we can connect two distant nodes $A$ and $B$ by finding a swapping route, i.e., a subset of nodes where swapping can be performed, that eventually connects $A$ and $B$. When all the states of the quantum network are initially maximally entangled, connecting any pair of nodes can be done optimally by just finding the minimum amount of swapping operations, as all swaps yield maximally entangled states. However, when the network is initially composed of non-maximally entangled states, swapping operations might not be sufficient to produce a maximally entangled state between any pair of nodes. Consequently, it becomes necessary to combine the swapping and distillation operations to achieve quantum percolation.

In this section, we describe the quantum percolation process analytically. We define an inequality that identifies the percolation threshold for a given number of LOCC between nodes at a short distance. These thresholds provide information on the amount of entanglement the network states should yield for the associated percolation to generate a maximally entangled state. Furthermore, we define a recursive relation that enables the computation of thresholds beyond the scope of LOCC combinations at a short distance.

\subsection{Local percolation strategies}

\begin{figure*}[t!]
    \centering
    \includegraphics[width=\linewidth]{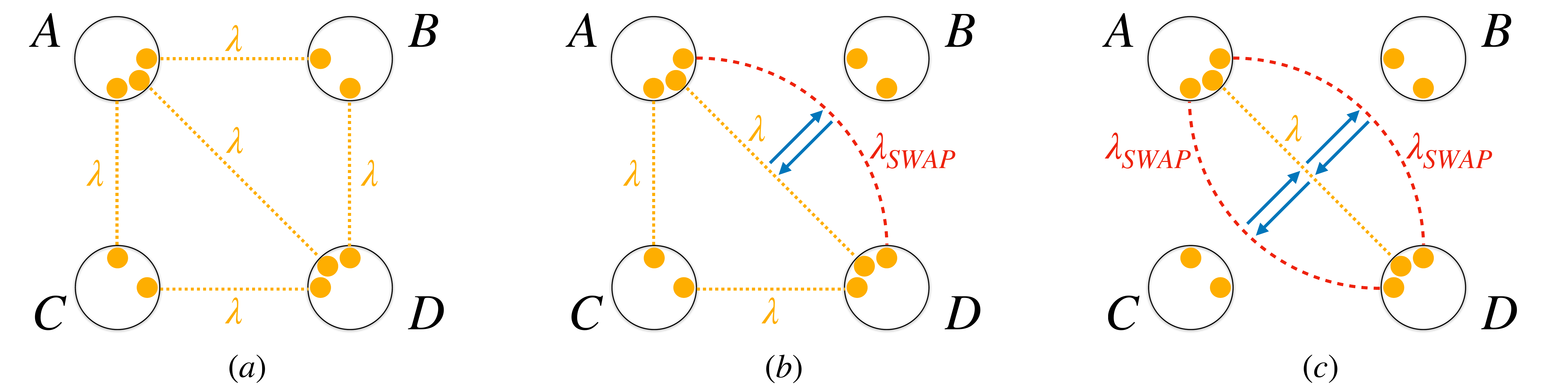}
    \caption{(a) 4-node quantum network arranged in a square. (b) Combined swapping and distillation to improve the entanglement of the state between $A$ and $D$. (c) Double swapping and distillation strategy to improve the entanglement of the state between $A$ and $D$.}
    \label{fig:square_ex}
\end{figure*}

Let us first define the distance between two nodes in a quantum network. Given any pair of nodes $A$ and $B$, if $n_s$ is the minimum number of swapping operations on the route to connect the pair, then $d = n_s + 1$ can be used as a measure of the distance between $A$ and $B$. Graphically, if each entangled state is represented as a link between two nodes, as in Figs.~\ref{fig:swapping} and~\ref{fig:distillation}, the distance $d$ between $A$ and $B$ is the minimum number of links that separate them. 

Let us now call a \textit{local} quantum percolation strategy a combination of operations that maximize the entanglement of the state between two nodes that can be connected with at most one swapping operation, that is, nodes at a distance of $1$ (nearest neighbors) or $2$ (at a distance of one swapping operation). Fig.~\ref{fig:square_ex} shows an example of a quantum network where 4 nodes are arranged in a square, with an additional link on the diagonal. The objective is to connect nodes $A$ and $D$ with a maximally entangled state. Assuming for simplicity that all states have the same Schmidt values and, consequently, equal entanglement, the strategy depends on the Schmidt coefficient $\lambda$. If $\lambda = \frac{1}{2}$, nothing needs to be done, as $A$ and $D$ are already connected. However, if $\lambda~>~\frac{1}{2}$, we might choose to perform distillation together with one (Fig.~\hyperref[fig:square_ex]{3b}) or two (Fig.~\hyperref[fig:square_ex]{3c}) swapping operations on alternative paths from $A$ to $D$. For instance, the Schmidt value $\lambda_{A\rightarrow D}$ of the state connecting $A$ to $D$ after one swapping and one distillation operations (Fig.\hyperref[fig:square_ex]{3b}) can be computed as:
\begin{equation}
    \lambda_{A\rightarrow D} = \max{\left(\frac{1}{2}, \lambda\cdot\lambda_{SWAP}\left(\lambda, \lambda\right)\right)}
    \label{eq:1S,2D}
\end{equation}
Solving the following inequality:
\begin{equation}
\label{eq:ineq_S,D}
    \lambda\cdot\lambda_{SWAP}\left(\lambda, \lambda\right) \leq \frac{1}{2}
\end{equation}
yields the threshold to produce a maximally entangled state with a local quantum percolation strategy involving one swapping operation and two distilled states.

We can now generalize Eq.~\ref{eq:ineq_S,D} to an arbitrary number of swaps and distillations. Let $\left\{1, 2, \dots, n\right\}$ be the set of positive integers smaller than or equal to $n$, and let $\lambda^{s\text{-}S, n\text{-}D}$ be the Schmidt value associated with the state that results from performing $s$ independent swapping operations and distilling $n$ states. Define $\mathbf{S}$ as the subset of length $s$ of photon pairs where a swapping operation is performed, and $\mathbf{K}$ as the subset of length $k = n - s$ of states that connect two nodes where the local quantum percolation strategy is performed. If the nodes are at distance $2$, then the set $\mathbf{K}$ is empty. Then, we can express the Schmidt value $\lambda^{s\text{-}S, n\text{-}D}$ resulting from any local percolation strategy as follows:

\small
\begin{align}
    \label{eq:lambdaSD}
    \lambda^{s\text{-}S, n\text{-}D} &= \displaystyle \max \left(\frac{1}{2}, \prod_{\left(j_1, j_2\right) \in \mathbf{S}} \lambda_{SWAP}\left(\lambda_{j_1}, \lambda_{j_2}\right) \cdot \prod_{i \in \mathbf{K}} \lambda_{i}\right)
\end{align}
\normalsize
Eq.~\ref{eq:ineq_S,D} directly generalizes as follows:

\small
\begin{equation}
    \displaystyle \prod_{\left(j_1, j_2\right) \in \mathbf{S}} \lambda_{SWAP}\left(\lambda_{j_1}, \lambda_{j_2}\right) \cdot \prod_{i \in \mathbf{K}} \lambda_{i} \leq \frac{1}{2}
    \label{eq:char_eq}
\end{equation}
\normalsize
Solving the above inequality for $\lambda_i~=~\lambda_{j_1}~=~\lambda_{j_2}~=~\lambda$, $i~\in~\mathbf{S}, j_1, j_2 \in \mathbf{K}$ yields the threshold $\lambda_{th}^{s\text{-}S, n\text{-}D}$ to connect any node pair within distance $2$ in a quantum network with a maximally entangled state. We call the quantity $\lambda_{th}^{s\text{-}S, n\text{-}D}$ the ``Schmidt value threshold'' of a local quantum percolation strategy involving $s$ swaps and $n$ distilled states. As an example, within our new formalism, the quantity $\lambda_{A\rightarrow D}$ from Eq.~\ref{eq:1S,2D} can be relabeled $\lambda^{1S,2D}$, representing the Schmidt value associated to a local strategy involving one entanglement swapping operation followed by distillation between two states.

\subsection{Connecting nodes at a large distance}

Our analytical framework allows us to compute percolation thresholds for local quantum percolation strategies that involve nodes at a maximum distance of two. By identifying a path of local strategies, we can connect any pair of nodes $A$ and $B$ at an arbitrary distance greater than two by swapping all the states resulting from the local combined operations:
\begin{equation}
    \lambda_{SWAP}\left(\lambda^{s_1\text{-}S,n_1\text{-}D}, \lambda_{SWAP}\left(\lambda^{s_2\text{-}S,n_2\text{-}D}, \dots\right)\right)
\end{equation}
If such a path exists, the percolation thresholds for local strategies naturally extend to long-range percolation, as the final threshold can be estimated by taking the minimum of the local thresholds:
\begin{equation}
    \lambda_{th} = \min\left(\lambda_{th}^{s_1\text{-}S,n_1\text{-}D}, \lambda_{th}^{s_2\text{-}S,n_2\text{-}D}, \dots\right)
\end{equation}

This generalization is possible because all local strategies lie along the same path. However, there exist more percolation thresholds that cannot be estimated with this description, as they originate from non-local percolation strategies that involve multiple swapping operations along a single path without intermediate distillation. Although estimating these new thresholds analytically is challenging, it is possible to extend the inequality from Eq.~\ref{eq:char_eq} to longer distances by recursively combining local percolation strategies. Let us define $\mathbf{\Lambda}^{(\tau)}$ as the set of Schmidt values associated with the states of the network after applying $\tau$ local percolation strategies. The set $\mathbf{\Lambda}^{(0)}$ is the set of initial Schmidt values, while all the Schmidt values of the network after multiple independent local percolation strategies belong to the set $\mathbf{\Lambda}^{(1)}$. Equivalently, we generalize the subsets $\mathbf{S}^{(\tau)}$ to represent the set of states involved in swapping at step $\tau$, and $\mathbf{K}^{(\tau)}$ to denote the set of states connecting the two nodes involved in the local percolation strategy at step $\tau$. In general, we can describe any Schmidt value belonging to the set $\mathbf{\Lambda}^{(\tau)}$ at step $\tau > 0$, denoted as $\lambda^{(\tau)}_{l}$, as the result of the combined LOCC on the Schmidt values belonging to the set $\mathbf{\Lambda}^{(\tau-1)}$:

\small
\begin{align}
    \lambda_{l}^{(\tau)} = &\displaystyle \max \left(\frac{1}{2}, \hspace{-2mm} \prod_{\left(j_1, j_2\right) \in \mathbf{S^{(\tau)}}} \hspace{-3.7mm} \lambda_{SWAP}\left(\lambda_{j_1}^{(\tau-1)}, \lambda_{j_2}^{(\tau-1)}\right) \cdot \hspace{-2mm} \prod_{i \in \mathbf{K^{(\tau)}}} \hspace{-1.3mm} \lambda_{i}^{(\tau-1)}\right),\\
    &\forall \ l = 1, \dots, \lvert\mathbf{\Lambda}^{(\tau)}\rvert \nonumber
\end{align}
\normalsize
Using the above expression, we can write a more general form for the inequality from Eq.~\ref{eq:char_eq}, taking into account the previous $\tau-1$ local percolation steps:

\small
\begin{equation}
    \displaystyle \prod_{\left(j_1, j_2\right) \in \mathbf{S^{(\tau)}}} \lambda_{SWAP}\left(\lambda_{j_1}^{(\tau-1)}, \lambda_{j_2}^{(\tau-1)}\right) \cdot \prod_{i \in \mathbf{K^{(\tau)}}} \lambda_{i}^{(\tau-1)} \leq \frac{1}{2}
    \label{eq:char_eq_recursive}
\end{equation}
\normalsize
By reconstructing the original expressions for $\lambda_l^{(\tau-1)}$ up to the recursive step 0, we can use this inequality to compute the percolation thresholds for pairs of nodes at a maximum distance of $\tau+1$. 

As an example, consider the following expression for a Schmidt value after a local strategy at step 2:

\small
\begin{equation}
    \lambda^{(2)}_l = \max\left(\dfrac{1}{2}, \lambda_{SWAP}\left(\lambda^{(1)}_{i_1}, \lambda^{(1)}_{j_1}\right)\cdot\lambda_{SWAP}\left(\lambda^{(1)}_{i_2}, \lambda^{(1)}_{j_2}\right)\right)
\end{equation}
\normalsize
For simplicity, suppose all the initial Schmidt values of the network are equal: $\lambda^{(0)}_{i} = \lambda \ \forall i$. Assuming $\lambda^{(1)}_{i_1}~=~\lambda_{i_2}^{(1)}~=~\lambda$ and $\lambda^{(1)}_{j_1} = \lambda^{(1)}_{j_2} = \lambda_{SWAP}\left(\lambda, \lambda\right)$, the expression can be rewritten as:
\small
\begin{equation}
    \label{eq:2s+1ss,2d}
    \lambda^{(2)}_l = \max\left(\dfrac{1}{2}, \left(\lambda_{SWAP}\left(\lambda, \lambda_{SWAP}\left(\lambda, \lambda\right)\right)\right)^2\right)
\end{equation}
\normalsize
This final expression can be used to find a new percolation threshold for a strategy that involves two distilled states resulting from recursive swapping operations.

\section{Simulating quantum networks with physics-informed heuristics}
\label{sec:pi_heuristics}
The analytical description of quantum percolation is useful to identify the percolation threshold associated with any given percolation strategy. However, this framework is reliable only when the Schmidt values of all initial states in the network are the same. Consider now a quantum network in which entanglement is distributed randomly across the states, introducing a form of \textit{quenched disorder} -- that is, static randomness in the initial entanglement configuration. In the presence of such disorder, it is not clear whether the dynamics of the percolation thresholds for given strategies would be preserved. Appendix~\ref{app:deviations} analytically describes the effect of non-uniform entanglement distribution on swapping and distillation operations, but extending this description to nodes at a large distance is quite challenging. Additionally, not all percolation thresholds can be easily computed or detected analytically, especially for those that are associated with non-local percolation strategies. To address these limitations, the idea of this work is to develop a numerical scheme that applies local percolation strategies in physical accordance with the analytical description, while also being able to find percolation paths between nodes at a high distance with large entanglement deviations. The purpose of this section is to explain the physics-informed heuristics employed to model quantum networks and simulate the quantum percolation process. The implementation of our algorithm can be found in~\cite{QNetsCode}.

Given a pair of distant nodes $A$ and $B$, the objective of the algorithm is to find a sequence of entanglement manipulation operations that connect $A$ to $B$ with a maximally entangled state. Each iteration of the algorithm consists of two main steps:
\begin{enumerate}
    \item combine locally optimal quantum percolation strategies;
    \item improve each generated state by exploring alternative paths.
\end{enumerate}
The two steps are then repeated for multiple samples, where the heuristic parameters are varied to explore many different solutions. After completion of the sampling process, the algorithm ultimately chooses the solution that achieves the highest entanglement between $A$ and $B$, while minimizing the number of original states destroyed.

\subsection{Combining local quantum percolation strategies}
\label{sec:local_strat}
\begin{figure*}[ht!]
  \begin{subfigure}{0.3\textwidth}
      \caption{}
      \label{fig:psexample_a}
      \centering
      \includegraphics[width=5.5cm]{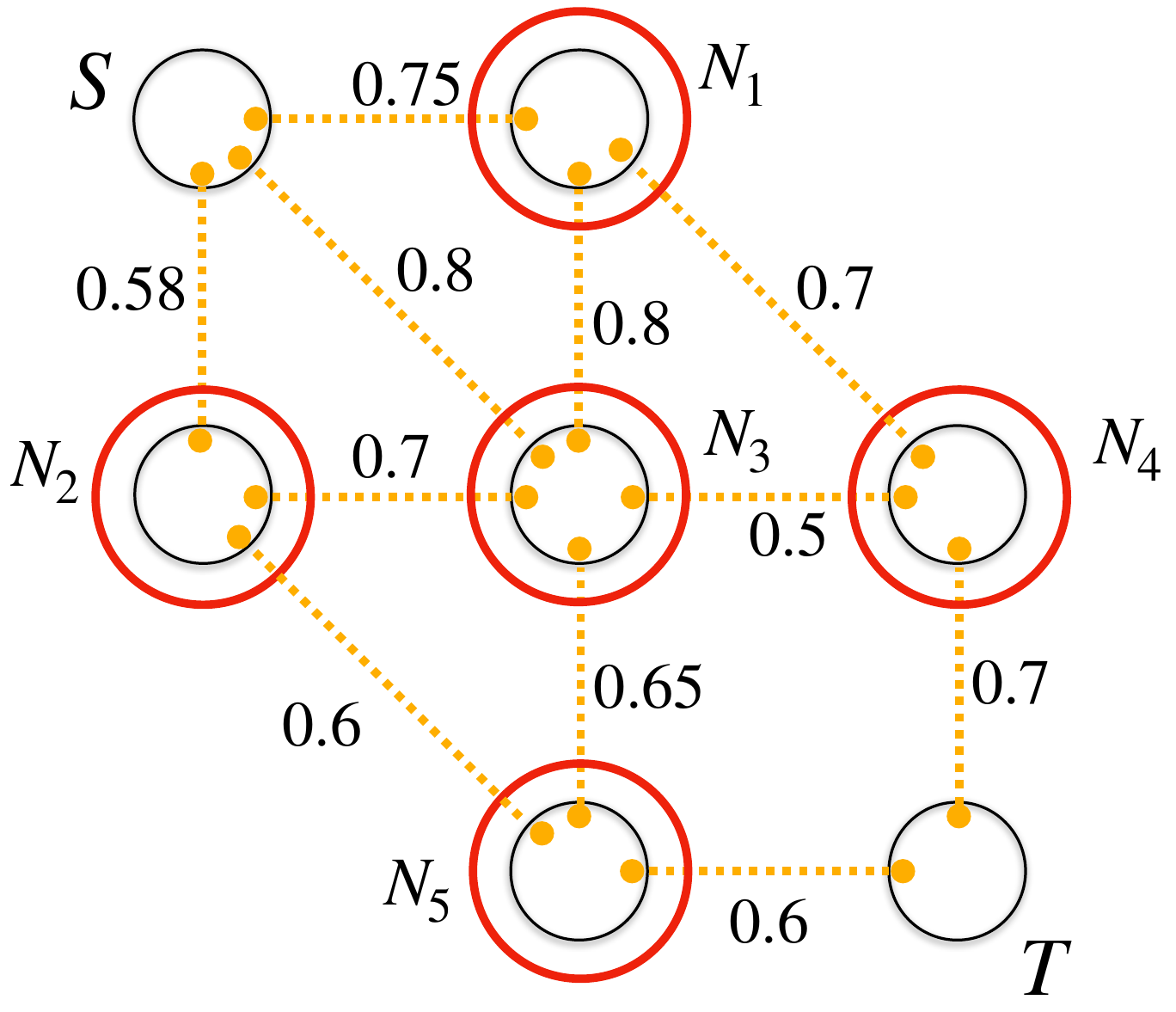}
  \end{subfigure}
  \hspace{1mm}
  \begin{subfigure}{0.3\textwidth}
      \caption{}
      \label{fig:psexample_b}
      \centering
      \includegraphics[width=5.5cm]{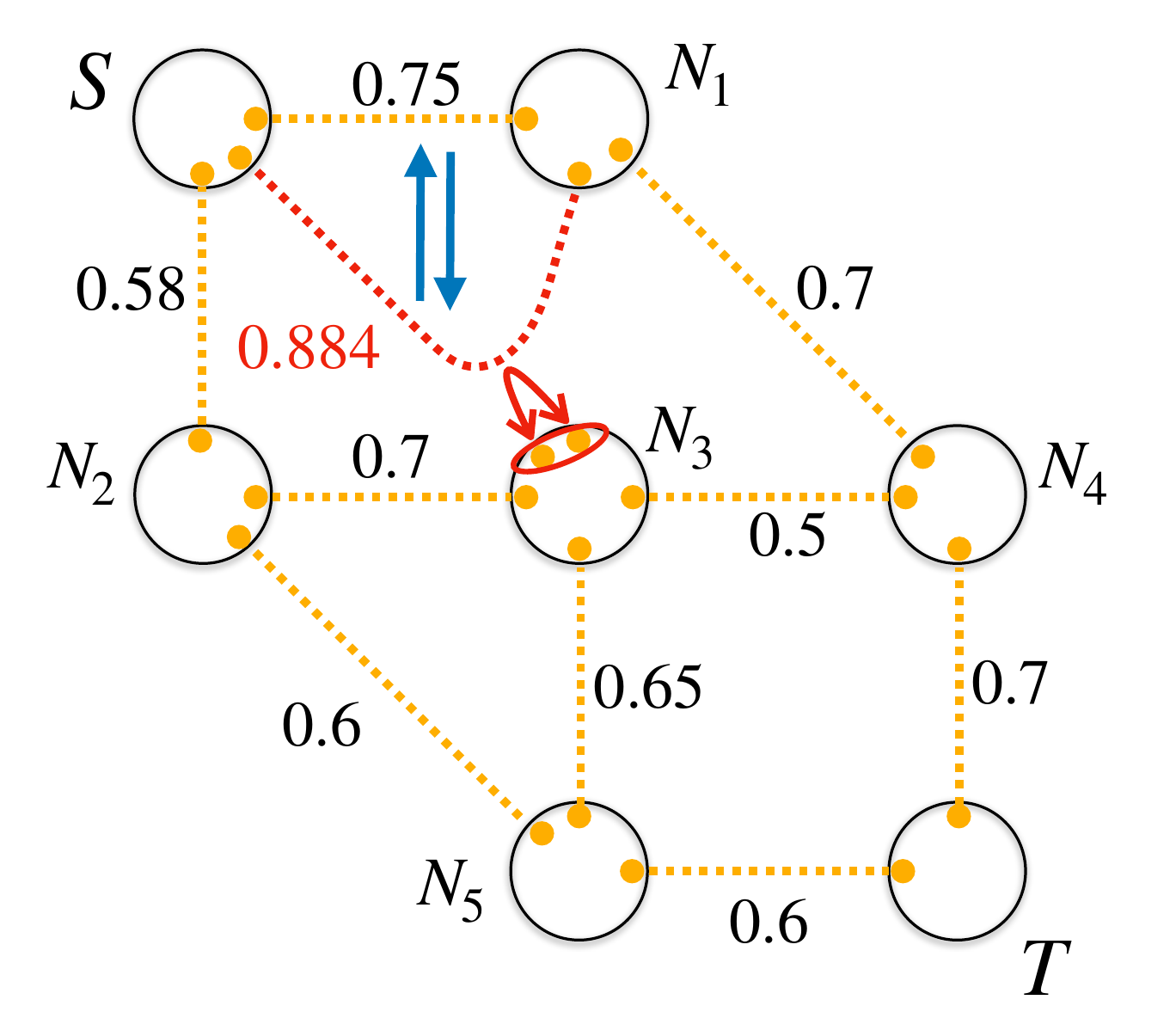}
  \end{subfigure}
  \hspace{1mm}
  \begin{subfigure}{0.3\textwidth}
      \caption{}
      \label{fig:psexample_c}
      \centering
      \includegraphics[width=5.5cm]{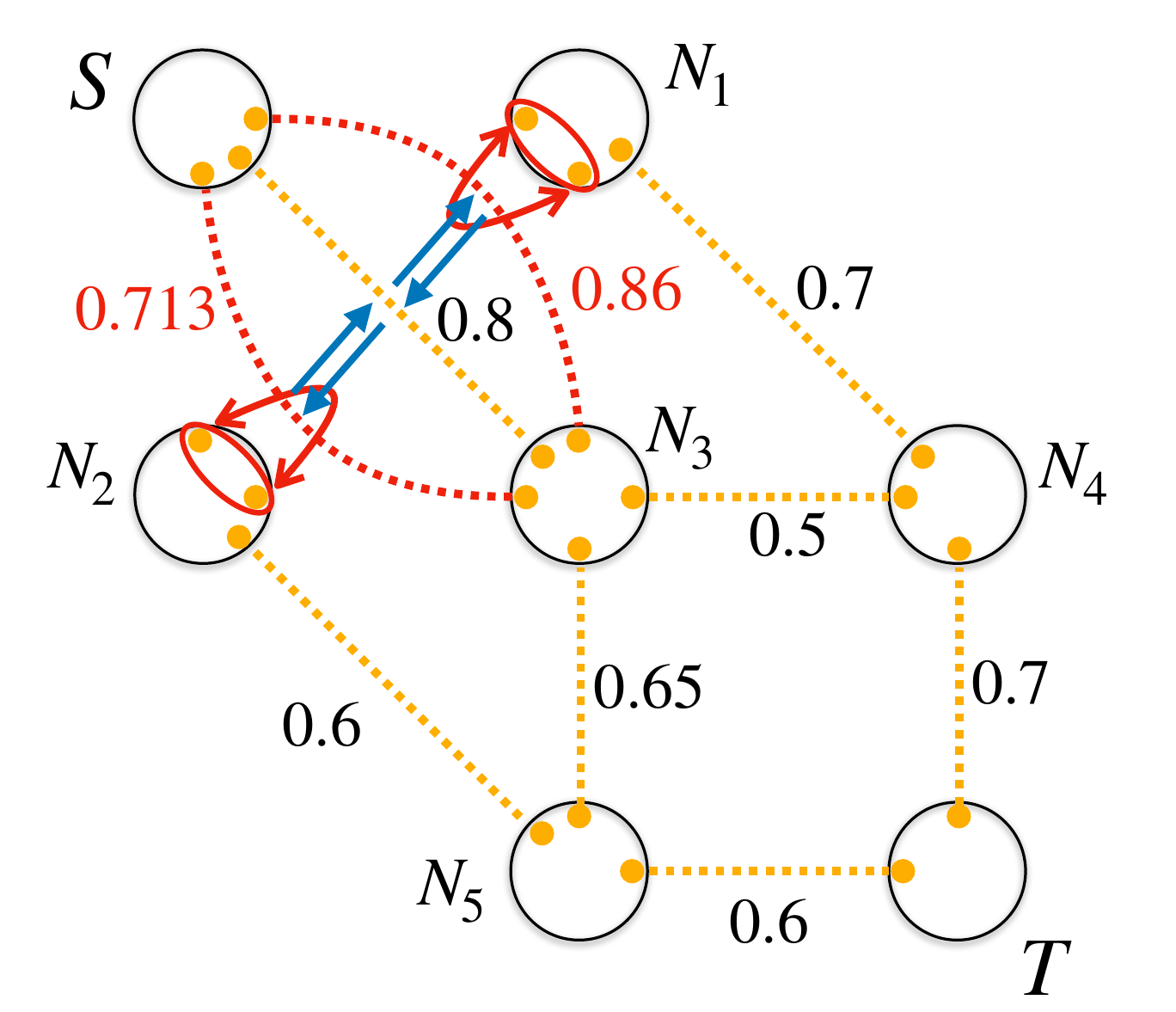}
  \end{subfigure}
  \hspace{1mm}
  \begin{subfigure}{0.3\textwidth}
      \caption{}
      \label{fig:psexample_d}
      \centering
      \includegraphics[width=5.5cm]{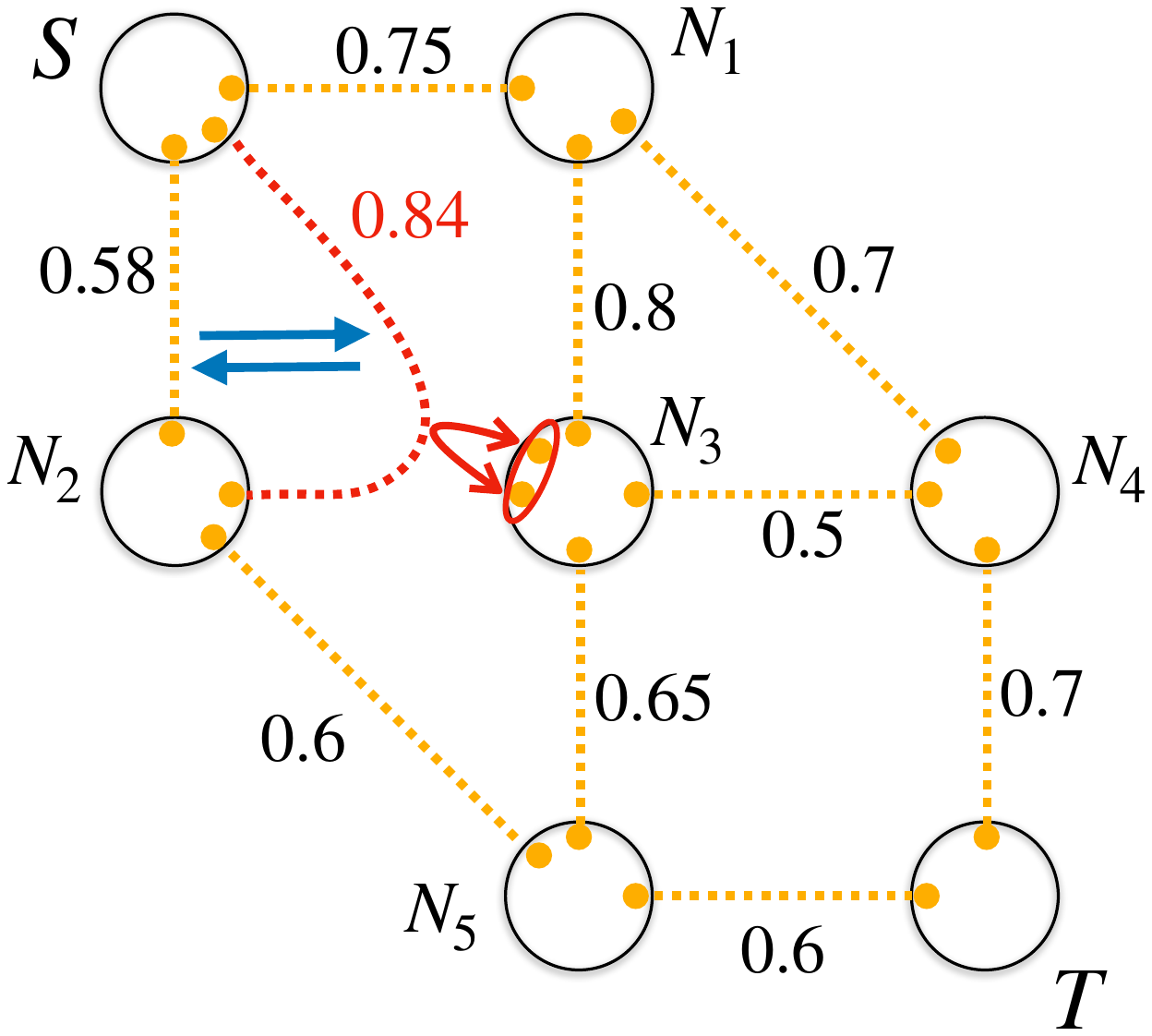}
  \end{subfigure}
  \hspace{1mm}
  \begin{subfigure}{0.3\textwidth}
      \caption{}
      \label{fig:psexample_e}
      \centering
      \includegraphics[width=5.5cm]{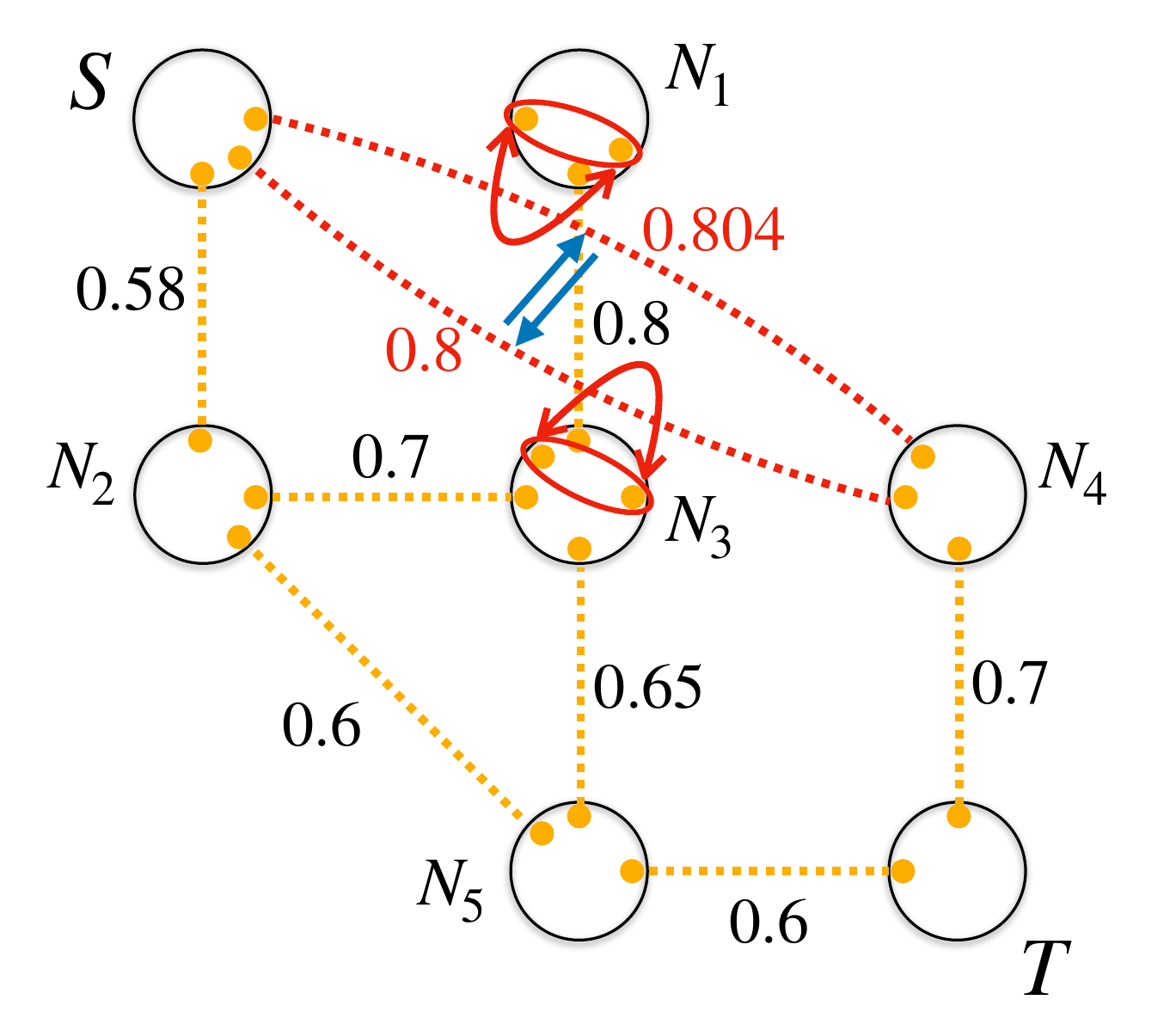}
  \end{subfigure}
  \hspace{1mm}
  \begin{subfigure}{0.3\textwidth}
      \caption{}
      \label{fig:psexample_f}
      \centering
      \includegraphics[width=5.5cm]{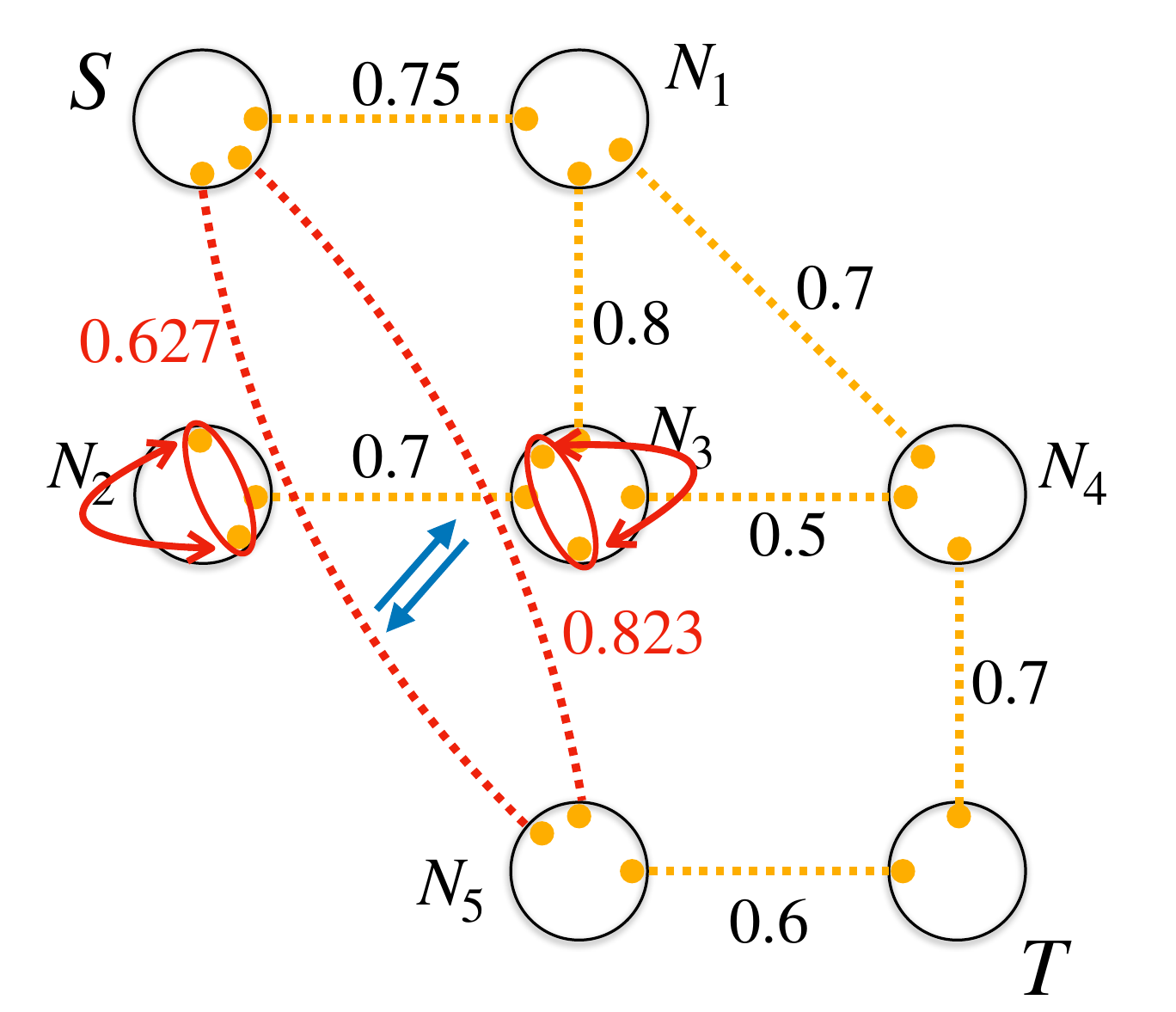}
  \end{subfigure}
  \caption{Schematic of the iteration process of the local-strategy-finding routine for a quantum network with 7 nodes. \hyperref[fig:psexample_a]{(a)}~The objective of the algorithm is to connect the starting node $S$ to the final target node $T$. In the first step, the algorithm selects all nodes at a distance of at most two from $S$. In this case, it finds five nodes that satisfy this constraint and iterates over each of them to find an optimal local percolation strategy to connect them to $S$. \hyperref[fig:psexample_b]{(b)}~$N_1$ is a neighbor of $S$, but their shared state is not maximally entangled. Therefore, the optimal local quantum percolation strategy between the two nodes involves an additional swapping operation on node $N_3$. The final state yields Schmidt value $\lambda_{S, N_1} = \max\left(0.5, 0.75\cdot\lambda_{SWAP}\left(0.8, 0.8\right)\right) \approx 0.663$. \hyperref[fig:psexample_c]{(c)}~The optimal local quantum percolation strategy from $S$ to $N_3$ involves distillation between the original shared state between $S$ and $N_3$ and two other states resulting from swapping operations on $N_1$ and $N_2$. The final state yields Schmidt value $\lambda_{S, N_3} = \max\left(0.5, 0.8\cdot\lambda_{SWAP}\left(0.75, 0.8\right)\cdot\lambda_{SWAP}\left(0.58, 0.7\right)\right) = 0.5$. While the final state is maximally entangled, it was necessary to destroy 5 original states of the network to generate it. \hyperref[fig:psexample_d]{(d)}~The optimal local quantum percolation strategy between $S$ and $N_2$ involves the distillation between the original shared state between $S$ and $N_2$ and another state resulting from a swapping operation on $N_3$. The final state yields Schmidt value $\lambda_{S, N_2} = \max\left(0.5, 0.58\cdot\lambda_{SWAP}\left(0.8, 0.7\right)\right) = 0.5$, meaning this local quantum percolation strategy creates a maximally entangled state by destroying 3 original states, less than in the percolation to $N_3$. \hyperref[fig:psexample_e]{(e)}-\hyperref[fig:psexample_f]{(f)}~The local quantum percolation strategies to $N_4$ and $N_5$ both involve the distillation of two states resulting from swapping operations, yielding states with Schmidt values $\lambda_{S, N_4} = \max\left(0.5, \lambda_{SWAP}\left(0.75, 0.7\right)\cdot\lambda_{SWAP}\left(0.8, 0.5\right)\right) \approx 0.6432$ and $\lambda_{S, N_5} = \max\left(0.5, \lambda_{SWAP}\left(0.58, 0.6\right)\cdot\lambda_{SWAP}\left(0.8, 0.65\right)\right) \approx 0.516$, respectively. As neither of them yields a maximally entangled state while destroying 4 original states of the network, the local percolation strategy between $S$ and $N_2$ is considered the optimal solution at this step.}
    \label{fig:percol_strat_example}
\end{figure*}
In the first part of the algorithm, the goal is to create a link between $A$ and $B$ by swapping multiple states resulting from independent local percolation strategies. To achieve this, we first define a routine that selects an intermediate node to draw a starting node closer to a final target node. This is accomplished by finding a local quantum percolation strategy that connects the starting node and the intermediate node to a state that has the maximum possible entanglement. 

\begin{figure*}[ht!]
  \begin{subfigure}{0.23\textwidth}
      \caption{}
      \label{fig:diagonal_square_lattice_b}
      \centering
      \includegraphics[width=4cm]{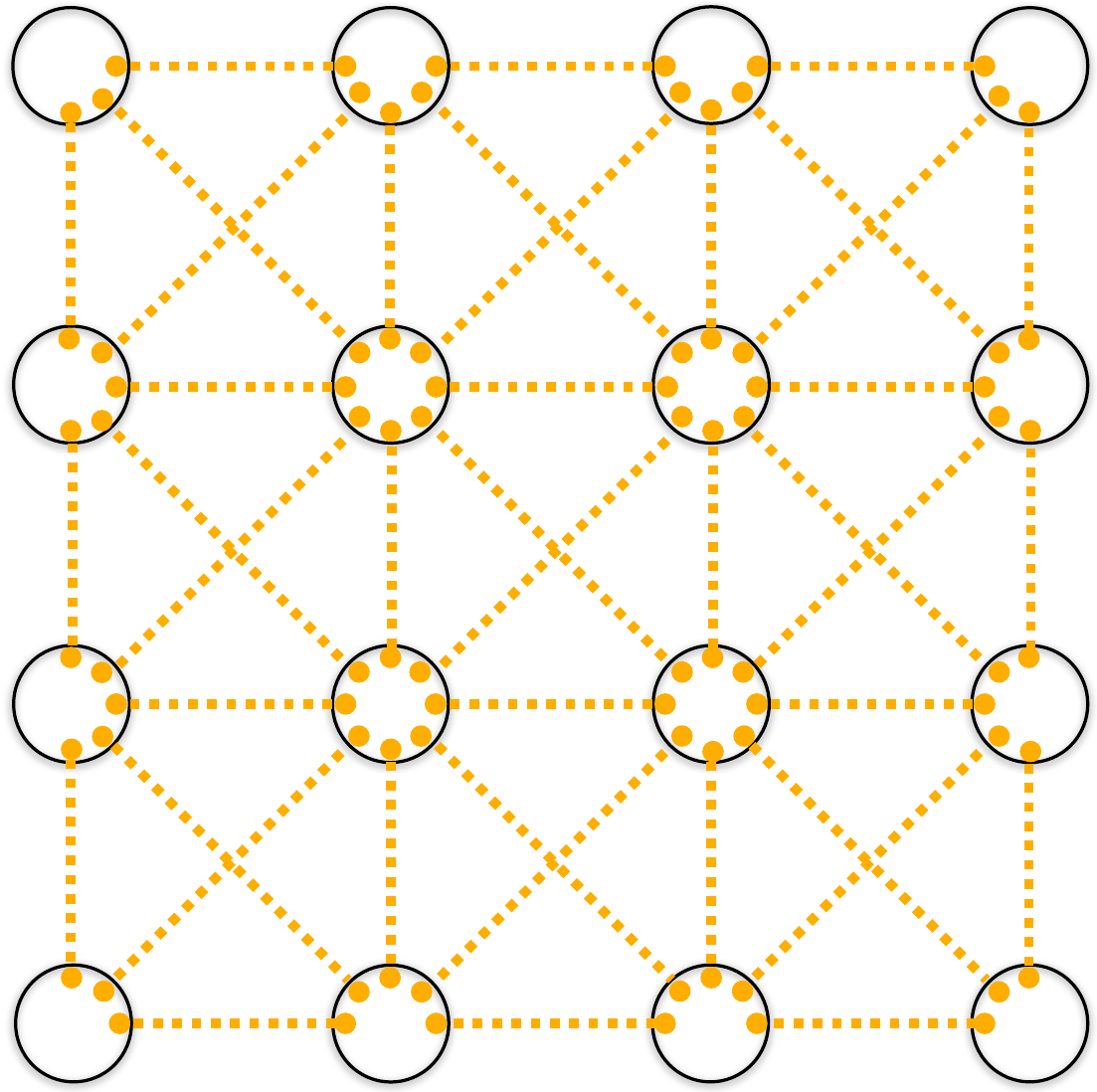}
  \end{subfigure}
  \hspace{1mm}
  \begin{subfigure}{0.23\textwidth}
      \caption{}
      \label{fig:fully_connected_honeycomb_d}
      \centering
      \includegraphics[width=4cm]{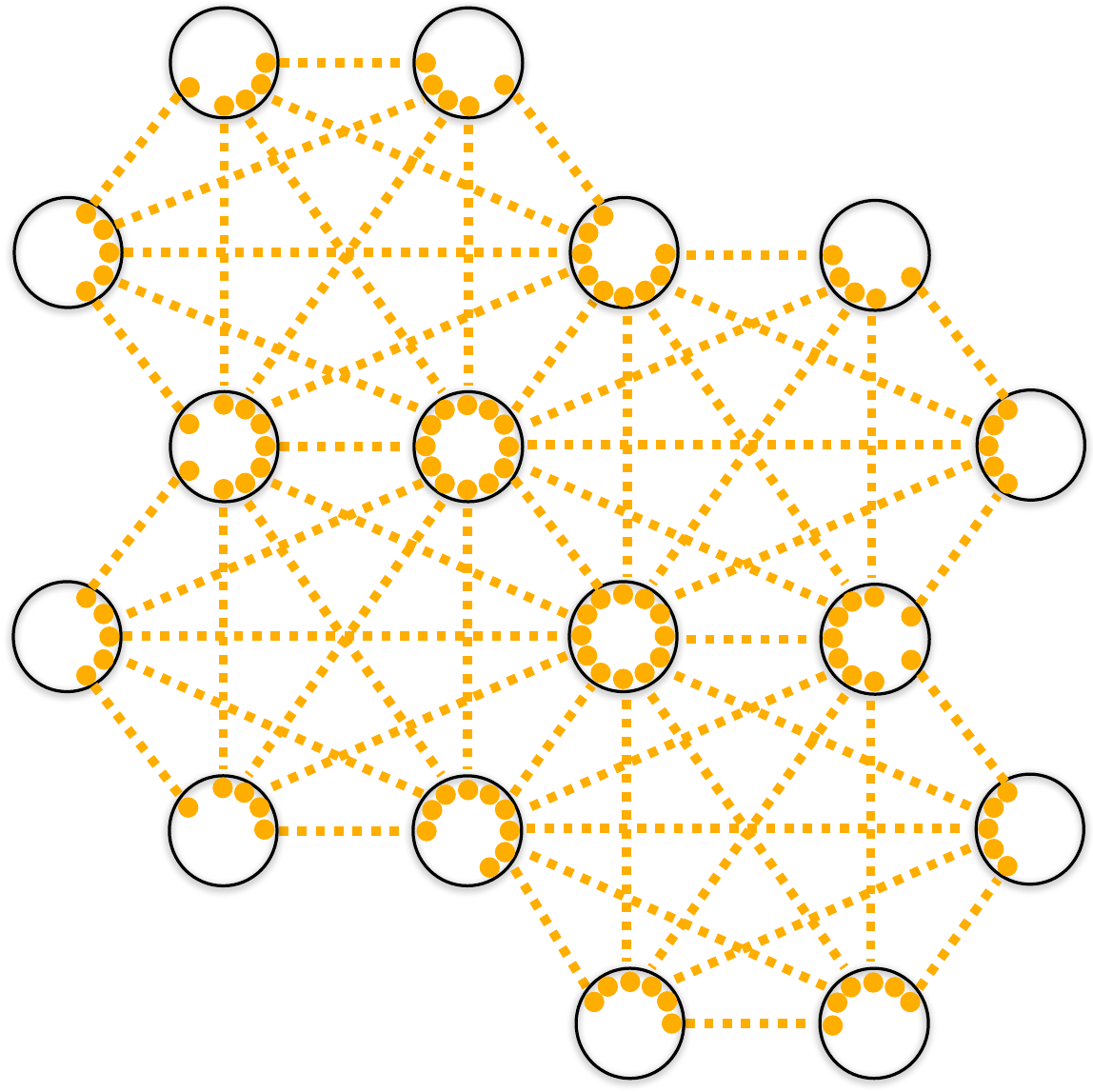}
  \end{subfigure}
  \hspace{1mm}
  \begin{subfigure}{0.23\textwidth}
      \caption{}
      \label{fig:square_lattice_a}
      \centering
      \includegraphics[width=4cm]{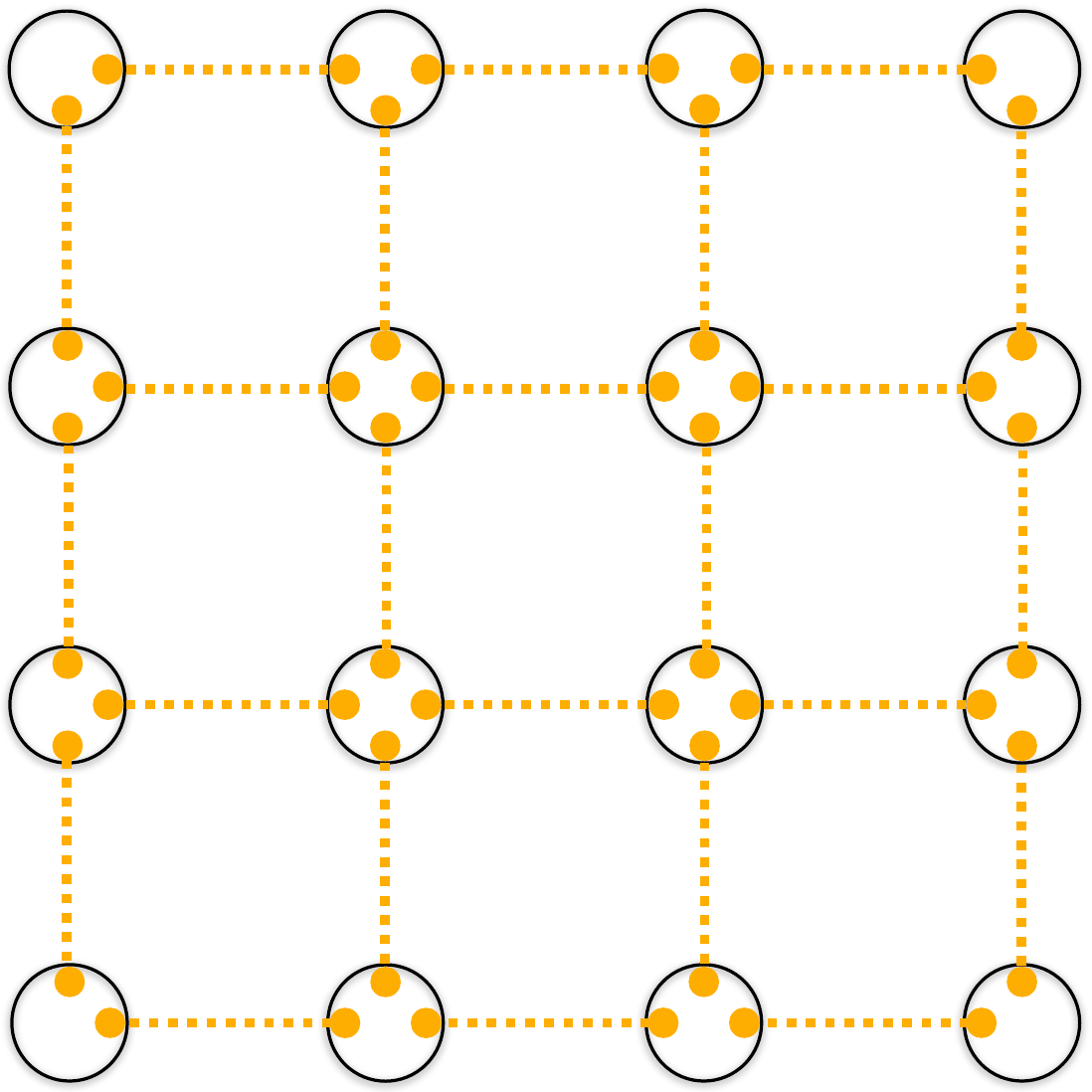}
  \end{subfigure}
  \hspace{1mm}
  \begin{subfigure}{0.23\textwidth}
      \caption{}
      \label{fig:honeycomb_c}
      \centering
      \includegraphics[width=4cm]{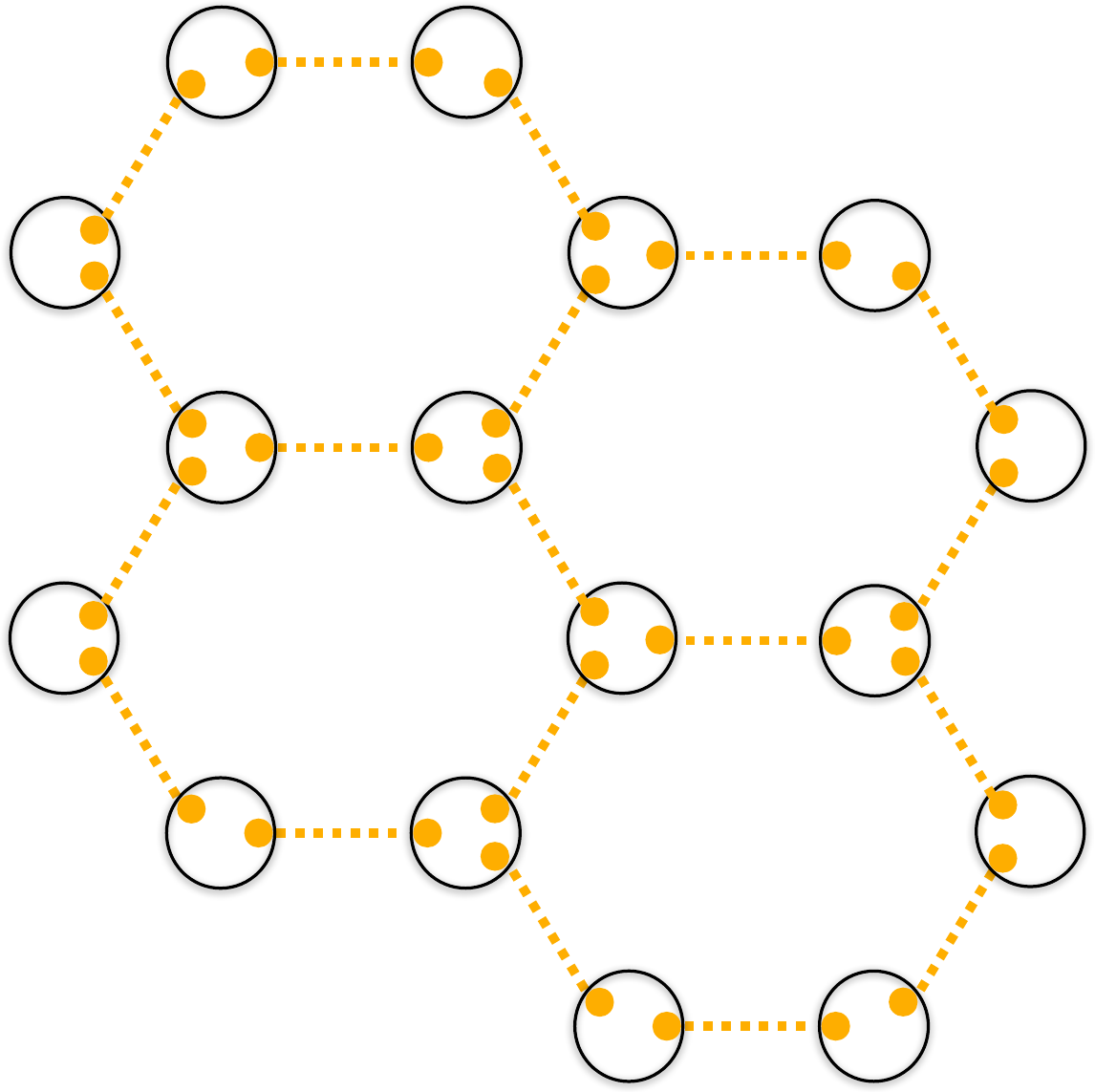}
  \end{subfigure}
  \caption{Example depictions of tested quantum network topologies. \hyperref[fig:diagonal_square_lattice_b]{(a)}~4x4 diagonal square lattice topology. The unit cells of this topology are fully connected squares, with additional states on the diagonals. \hyperref[fig:fully_connected_honeycomb_d]{(b)}~Fully connected honeycomb lattice topology with 4 fully connected hexagons. \hyperref[fig:square_lattice_a]{(c)}~4x4 square lattice topology. \hyperref[fig:honeycomb_c]{(d)}~Honeycomb lattice topology with 4 hexagons.}
    \label{fig:topologies}
\end{figure*}

This procedure starts by investigating potential intermediate nodes selected from the set of all nearest neighbors and all nodes located at distance two from the starting node. Initially, the chosen nodes are only those strictly closer to the final destination, meaning that the minimum number of swaps required to connect them to the final node is smaller than the number required from the initial node. For each node in this smaller set, the algorithm evaluates the minimum amount of swapping and distillation operations needed to generate a state with the highest possible entanglement between the starting node and the intermediate node. As explained in Sect.~\ref{sec:analytics}, percolation between two nodes at a distance of at most two can be effectively described by a local quantum percolation strategy, which yields a new state with Schmidt value $\lambda^{s\text{-}S,n\text{-}D}$, which is calculated directly using Eq.~\ref{eq:lambdaSD}. Since locally optimal solutions are the nodes that minimize both the Schmidt value and the amount of destroyed states, the algorithm can compute these quantities for all local percolation strategies between the initial node and the potential intermediate nodes. The resulting solutions are then compared using two cost functions: the Schmidt value and the number of destroyed states, with priority given to minimizing the Schmidt value. If multiple solutions of equal quality exist at any step, the algorithm can randomly select one from the set of viable candidates. 

Once an intermediate node has been chosen, the algorithm repeats this routine by setting the selected node as the new starting node, and keeps hopping to intermediate nodes until the final target node is reached. Fig.~\ref{fig:percol_strat_example} shows an example iteration of this algorithm. The full execution of the algorithm on the example network from Fig.~\ref{fig:percol_strat_example}, including later iterations of the local-strategy-finding routine, is detailed in Appendix~\ref{app:example_2}.

\subsection{Exploration of alternative paths and heuristics}
By combining locally optimal solutions with the approach introduced in the previous section, we obtain a greedy solution to the percolation problem. The resulting state between $A$ and $B$ is maximally entangled only if the states generated by all local percolation strategies are also maximally entangled:
$$\lambda_{A\rightarrow B} = \frac{1}{2} \iff \lambda^{s_1\text{-}S,n_1\text{-}D} = \dots = \lambda^{s_\tau\text{-}S,n_\tau\text{-}D} = \frac{1}{2}$$
If the final state is not maximally entangled, the protocol fails, and it is therefore necessary to explore alternative paths around the non-maximally entangled states in the path between $A$ and $B$.

To achieve this, the algorithm iterates over each non-maximally entangled state generated by local percolation strategies and exploits the routine from Sect.~\ref{sec:local_strat} to find an alternative path between the two nodes connected by the current state, using them as initial and final node of the routine. This search is repeated until the state, after distilling it with the newly found state(s), becomes maximally entangled. A maximum number of (10) iterations is also set in case the algorithm is unable to find a solution.

In addition to the alternative path search, the algorithm takes into account multiple heuristics to explore local solutions and expand the range of possible paths from any pair of nodes. By default, the local strategy-finding routine searches for optimal solutions, excluding those that either fail to produce strictly maximally entangled states or destroy more states than the optimal solutions. However, a locally optimal solution is not necessarily globally optimal. As shown in the example provided in Appendix~\ref{app:example_2}, a solution initially ignored for being locally suboptimal can actually lead to the globally optimal entanglement percolation path. For this reason, we introduce a parameter that gradually relaxes the constraints on the local optimality of the solution over multiple samples, increasing the set of possible local solution that are considered ``valid'' in each step. Additionally, as the sampling continues, we relax the constraint that intermediate nodes must strictly reduce their distance to the target node. This allows the algorithm to explore paths involving nodes at equal or even greater distances from the target, potentially discovering nontrivial paths that lead to higher entanglement between the source and target. 

By sampling multiple solutions of the procedure while varying the parameters that define the heuristic, the algorithm generates multiple solutions to percolate a quantum network to connect any pair of distant nodes. Finally, if any pair of these solutions are suboptimal but mutually independent -- meaning their paths do not cross -- the algorithm considers the distillation between the two generated states as an additional solution to increase entanglement in the final state.

\begin{figure*}[ht!]
  \begin{subfigure}{0.45\textwidth}
      \label{fig:diagonal_square_lattice_plot_a}
      \centering
      \includegraphics[width=8.2cm]{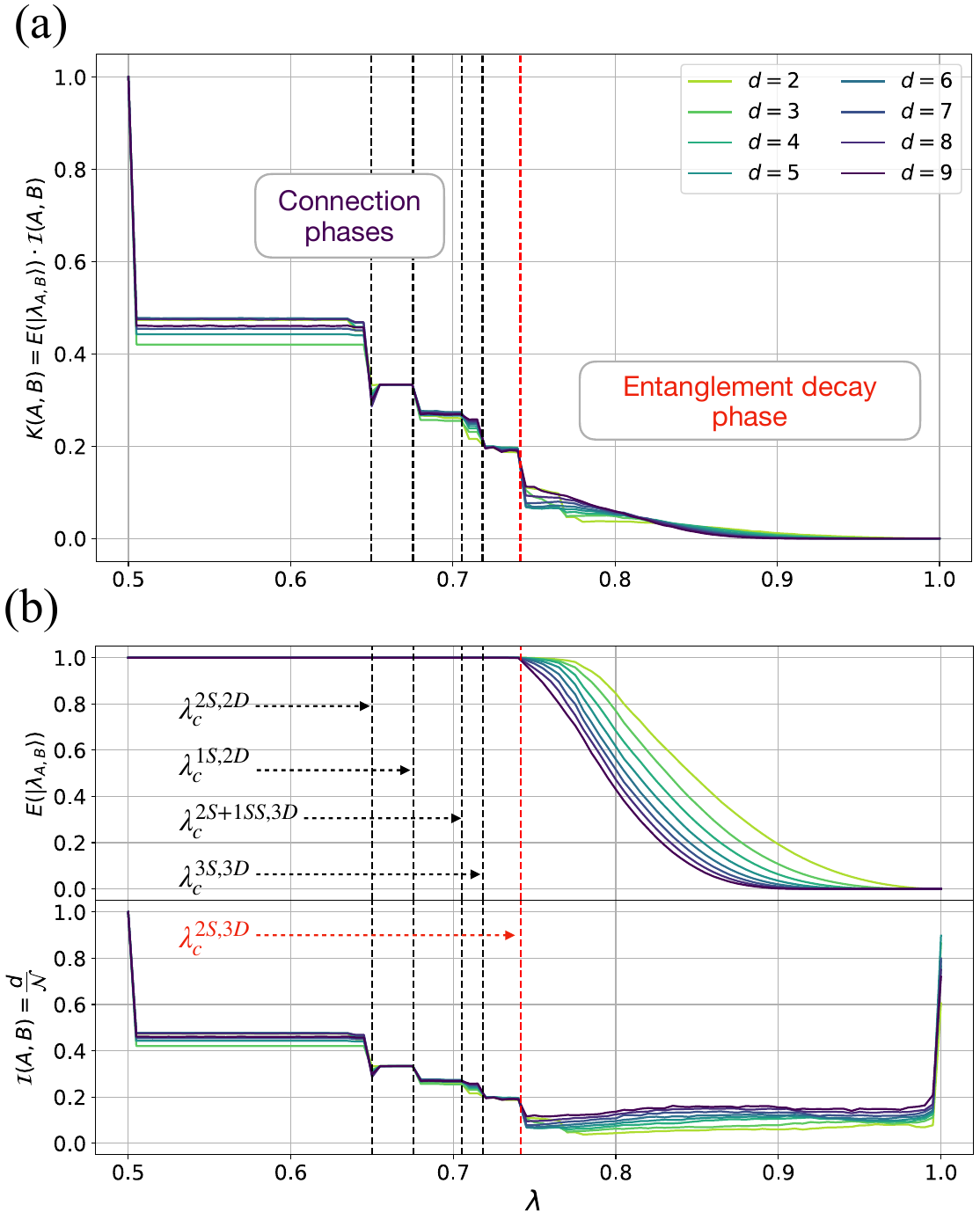}
  \end{subfigure}
  \hspace{5mm}
  \begin{subfigure}{0.45\textwidth}
      \label{fig:diagonal_square_lattice_plot_b}
      \centering
      \includegraphics[width=8.2cm]{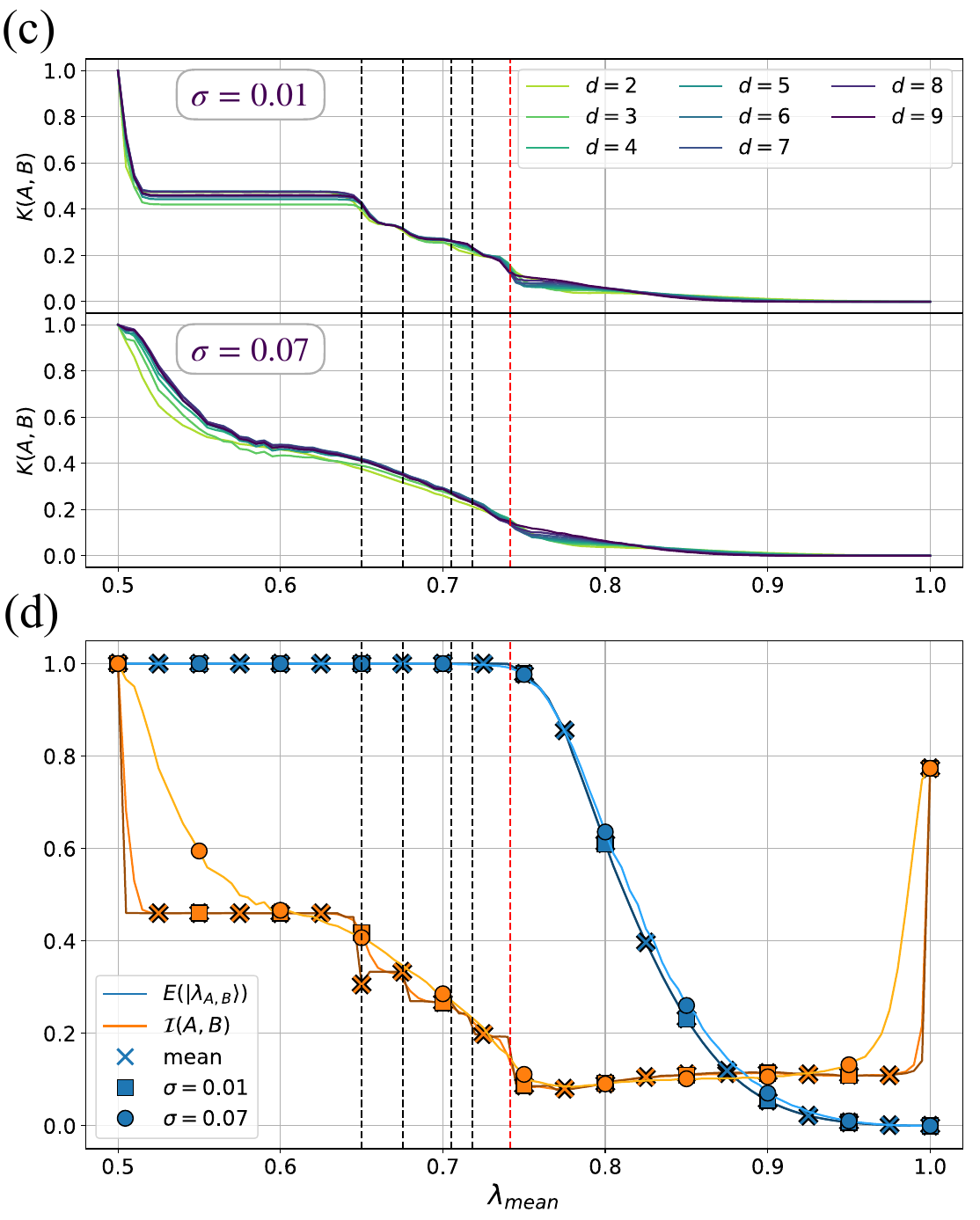}
  \end{subfigure}
  \caption{Results for a $10\times10$ diagonal square lattice quantum network. \hyperref[fig:diagonal_square_lattice_plot_a]{(a)}~Connectivity of the network after quantum percolation between all possible distant node pairs, plotted against the Schmidt value assigned to all the initial states of the network. Each curve corresponds to the average connectivity of all pairs of nodes at a distance $d$. The vertical dashed lines indicate the percolation thresholds predicted with the analytical description. \hyperref[fig:diagonal_square_lattice_plot_a]{(b)}~Entanglement of the final state between distant nodes $A$ and $B$ (top) and integrity of the network (bottom) after quantum percolation between all possible pairs at distance $d$, plotted against the Schmidt value assigned to all the initial states of the network. \hyperref[fig:diagonal_square_lattice_plot_b]{(c)}~Connectivity of the network after quantum percolation between all possible pairs at distance $d$, plotted against the mean Schmidt value of the initial states of the network. The initial Schmidt values are drawn from a truncated normal distribution, with an enforcement on the mean $\lambda_{mean}$, with standard deviations $\sigma = 0.01$ (top) and $\sigma = 0.07$ (bottom). The results are averaged over 10 different network samples. \hyperref[fig:diagonal_square_lattice_plot_b]{(d)}~Comparison of entanglement and integrity between quantum networks with initial equal Schmidt values (mean), and drawn from a truncated normal distribution with mean enforcement, with standard deviations $\sigma = 0.01, 0.07$.}
    \label{fig:diagonal_square_lattice_plots}
\end{figure*}

\section{Network integrity and connectivity}
\label{sec:connectivity}
Our analytical description of entanglement percolation reveals that the less entanglement is distributed across the initial quantum network, the greater the number of states required to generate maximally entangled states. Quantifying the disruption in the network in terms of the number of states destroyed during percolation can provide a basis for defining a measure of network resilience, as well as for evaluating the quality of a chosen percolation path. In an experimental setting, it is important to minimize the number of states destroyed during percolation, since repeating the procedure for a different pair of nodes requires the reconstruction of the original states. Furthermore, if a large number of the network's initial states need to be involved in LOCC to connect two distant nodes, the network becomes almost completely disconnected and no further communication among other nodes is possible. Therefore, when analyzing how quantum networks are affected by the routing process for two nodes $A$ and $B$, it is important to both maximize the entanglement of the final state between $A$ and $B$ (Eq.~\ref{eq:ent}) and minimize the amount of disruption caused by this process.

Let us then define a quantity for the network ``integrity'':
\begin{equation}
    \mathcal{I}(A,B) := \frac{d(A,B)}{\mathcal{N}(A,B)}
\end{equation}
where $\mathcal{N}(A,B)$ is the number of original states destroyed by the performed operations in the percolation between nodes $A$ and $B$ of the network, while $d(A,B)$ is the distance between $A$ and $B$, defined earlier in Sect.~\ref{sec:qnets_gen} as the minimum number of states separating $A$ and $B$. If the integrity falls below one, then the number of states destroyed by the percolation strategy increases more quickly than its theoretical minimum values, corresponding to the distance between $A$ and $B$.

In the percolation of a quantum network model with non-maximally entangled states, the integrity $\mathcal{I}(A,B)$ can be used as an indicator of how much destruction the network has to undergo to connect $A$ and $B$ with a maximally entangled state.  If the network is initially perfect, that is, all states have maximum entanglement $1$, swapping the minimum number of states will still produce a perfect state between any node $A$ and $B$, which means $\mathcal{I}(A,B) = 1$. However, even a small amount of imperfection in the network's initial entanglement disallows the sole use of swapping to produce perfectly entangled states. The use of distillation in this case becomes necessary, at the cost of increasing the amount of destroyed states and, consequently, decreasing the integrity of the network. We now define the connectivity $K(A,B)$ to relate the entanglement of the final state between $A$ and $B$ and the integrity of the network at the end of the process:
\begin{equation}
    K(A, B) := E\left(\ket{\lambda_{AB}}\right)\cdot\mathcal{I}(A,B)
\end{equation}
We will use this parameter, together with entanglement and integrity, to evaluate each tested network topology based on how it gets affected by the routing process.

\section{Results}
\label{sec:results}

In this section, we outline the results obtained. We test our physics-informed heuristic algorithm on four different topologies:
\begin{enumerate}
    \item a $10 \times 10$ diagonal square lattice quantum network (Fig.~\ref{fig:diagonal_square_lattice_b});
    \item a honeycomb lattice with 36 fully connected hexagons (Fig.~\ref{fig:fully_connected_honeycomb_d});
    \item a $10 \times 10$ square lattice quantum network (Fig.~\ref{fig:square_lattice_a});
    \item a honeycomb lattice with 36 hexagons (Fig.~\ref{fig:honeycomb_c}).
\end{enumerate}
We simulate the percolation of a quantum network to connect a pair of nodes at distance $d$, and repeat this simulation for all possible pairs of nodes in the network, dividing node pairs by distance. To achieve a solution as close as possible to the optimal one, we sample 600 different percolation paths for each node pair. We first study quantum networks with all initial Schmidt values set at a fixed value $\lambda$. Eventually, we compare the results with those of quantum networks with different initial Schmidt values, drawn from a truncated normal distribution around an enforced mean $\lambda_{mean}$. We choose two different distributions, with standard deviations $\sigma = 0.01$ and $\sigma = 0.07$, and, for each of them, we sample 10 different quantum networks, averaging the results. In this section, we only outline the results for the diagonal square lattice and the fully connected honeycomb topologies, while those for the regular square lattice and the honeycomb lattice are outlined in Appendix~\ref{app:square_honeycomb}. 

Fig.~\hyperref[fig:diagonal_square_lattice_plot_a]{6a} shows the network connectivity of the $10\times10$ diagonal square lattice quantum network after quantum percolation, plotted against the Schmidt value $\lambda$ assigned to all initial states of the network. Each curve is assigned a distance $d$, as it represents the average connectivity after performing quantum percolation to connect all possible pairs of nodes at such a distance. 
In the depicted setup, we can clearly distinguish jumps in the connectivity, which are independent of the distance between nodes. These jumps signify a transition in the main employed local percolation strategy that allows the creation of a maximally entangled state. As the Schmidt value $\lambda$ increases, the cost of creating a maximally entangled state at large distances increases only when the previously employed strategy does not guarantee maximal entanglement. In contrast, for Schmidt values belonging to the same regime, the cost does not change, as the employed strategy is always the same. This causes the creation of effective regimes of connectivity for quantum percolation, separated by percolation thresholds, we term ``connection phases''. Throughout these phases, since it is possible to optimally percolate the network by combining percolation strategies, the final state between all pairs of nodes is predicted to be maximally entangled, as shown in Fig.~\hyperref[fig:diagonal_square_lattice_plot_a]{6b}. Moreover, by looking at the entanglement curve, we observe that the algorithm is capable of finding an optimal percolation path up to a certain amount of entanglement imperfection. After the Schmidt value $\lambda$ becomes large enough to cross this point, the entanglement of the final state generated by quantum percolation starts to decay and the connection between the two distant nodes becomes weaker. For this reason, we call this regime the ``entanglement decay'' phase.

Starting from the numerical results, we can detect the percolation strategies that define the various connection phases and compute their corresponding Schmidt value thresholds using Eq.~\ref{eq:char_eq}. The vertical dashed lines in each plot indicate the percolation thresholds of the strategies detected with the analytical description.
A summary of these strategies for the diagonal square lattice is provided below.
\begin{itemize}
    \item Two swapping operations + distillation of the two resulting states. This strategy yields the threshold $\lambda_{th}^{2S,2D}~\approx~0.6498$, solution of the inequality: 
    $$\left(\lambda_{SWAP}(\lambda, \lambda)\right)^2 \leq \frac{1}{2}$$
    \item One swapping operation + distillation of the resulting state with a previously existing state. This strategy yields the threshold $\lambda_{th}^{1S, 2D}~\approx~0.675$, solution of the inequality:
    $$\lambda_{SWAP}(\lambda, \lambda)\cdot\lambda \leq \frac{1}{2}$$
    \item Two swapping operations + a swapping operation between an initial state and a state obtained with another swapping operation + distillation of the three resulting states. This strategy yields the threshold $\lambda_{th}^{2S+1SS, 3D}~\approx~0.705$, solution of the inequality:
    $$\left(\lambda_{SWAP}\left(\lambda, \lambda\right)\right)^2 \cdot \lambda_{SWAP}\left(\lambda, \lambda_{SWAP}\left(\lambda, \lambda\right)\right) \leq \dfrac{1}{2}$$
    \item Three swapping operations + distillation of the three resulting states. This strategy yields the threshold $\lambda_{th}^{3S,3D}~\approx~0.718$, solution of the inequality: 
    $$\left(\lambda_{SWAP}(\lambda, \lambda)\right)^3 \leq \frac{1}{2}$$
    \item Two swapping operations + distillation of the resulting states with a previously existing state. This strategy yields the threshold $\lambda_{th}^{2S, 3D}~\approx~0.742$, solution of the inequality:
    $$\left(\lambda_{SWAP}(\lambda, \lambda)\right)^2\cdot\lambda \leq \frac{1}{2}$$
\end{itemize} 
The previously described jumps in the connectivity curve occur exactly at the analytically predicted percolation threshold, proving that the analytical description is consistent with the numerical results. We emphasize the presence of a jump at the threshold $\lambda_{th}^{2S+1SS,3D}$, which results from a non-local percolation strategy. This signifies that the algorithm, despite being based on the combination of local strategies, successfully detects a transition for a non-local percolation strategy, highlighting its effectiveness. Moreover, the percolation threshold $\lambda_{th}^{2S,3D}$ is highlighted in red in the figure because it is associated with the last possible local strategy for the diagonal square lattice topology. Since no more optimal percolation strategies are available after this point, it corresponds to the transition point between the connection phases and the entanglement decay phase.

\begin{figure*}[ht!]
  \begin{subfigure}{0.45\textwidth}
      \label{fig:fully_connected_honeycomb_a}
      \centering
      \includegraphics[width=8.2cm]{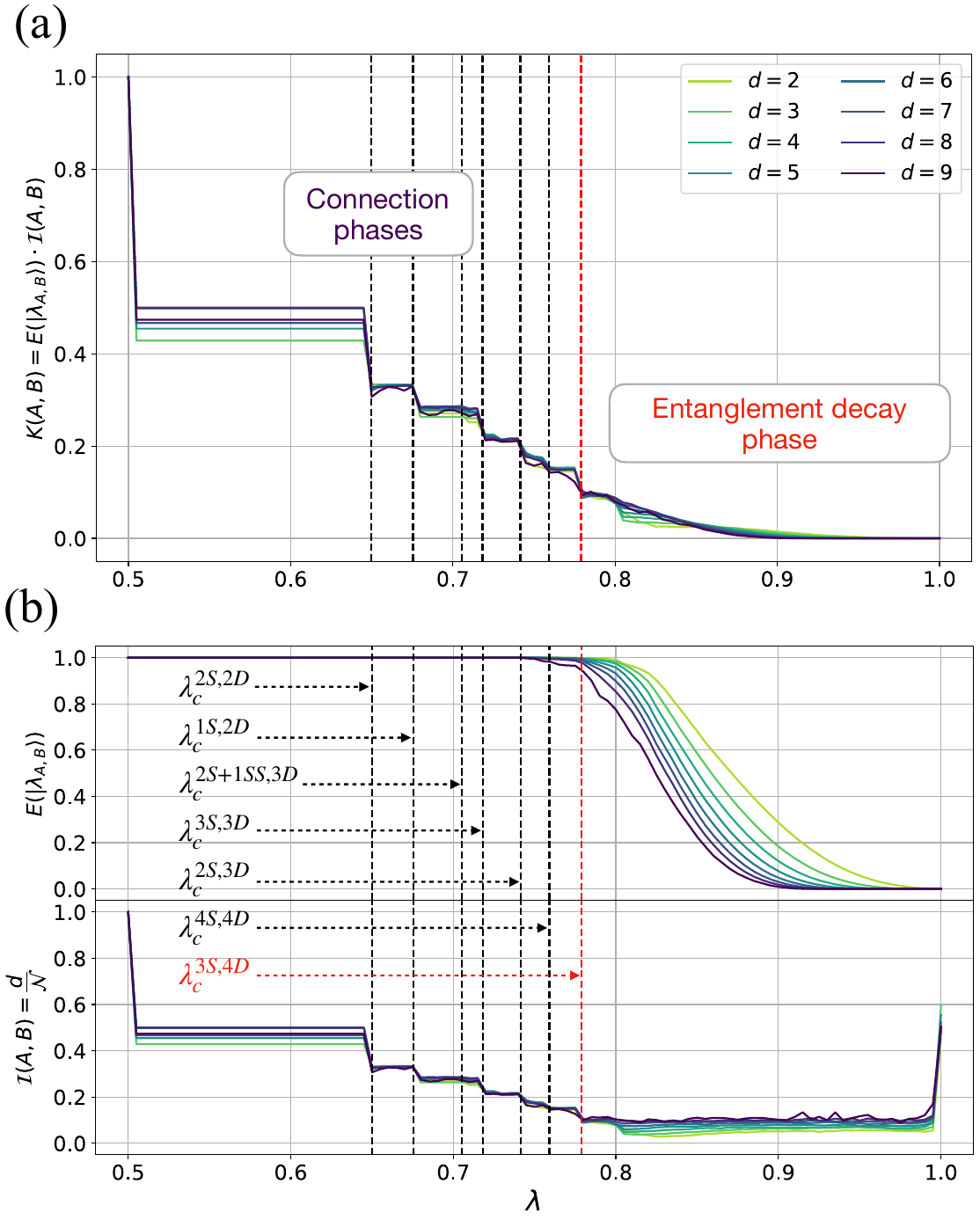}
  \end{subfigure}
  \hspace{5mm}
  \begin{subfigure}{0.45\textwidth}
      \label{fig:fully_connected_honeycomb_b}
      \centering
      \includegraphics[width=8.2cm]{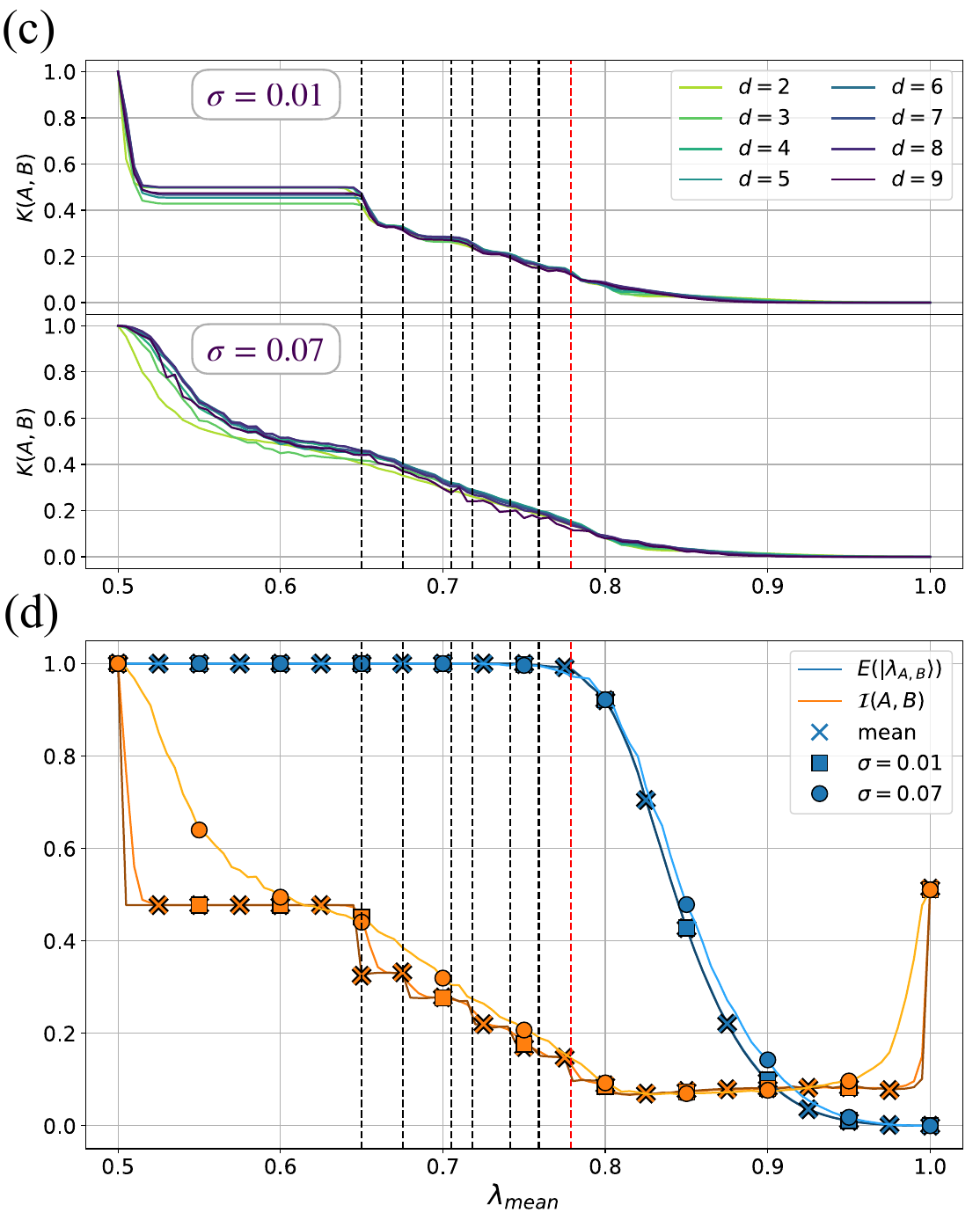}
  \end{subfigure}
  \caption{Results for a fully connected honeycomb lattice quantum network with 96 nodes arranged in 36 fully connected hexagons. \hyperref[fig:fully_connected_honeycomb_a]{(a)}~Connectivity of the network after quantum percolation between all possible distant node pairs, plotted against the Schmidt value assigned to all the initial states of the network. Each curve corresponds to the average connectivity of all pairs of nodes at a distance $d$. The vertical dashed lines indicate the percolation thresholds predicted with the analytical description. \hyperref[fig:fully_connected_honeycomb_a]{(b)}~Entanglement of the final state between distant nodes $A$ and $B$ (top) and integrity of the network (bottom) after quantum percolation between all possible pairs at distance $d$, plotted against the Schmidt value assigned to all the initial states of the network. \hyperref[fig:fully_connected_honeycomb_b]{(c)}~Connectivity of the network after quantum percolation between all possible pairs at distance $d$, plotted against the mean Schmidt value of the initial states of the network. The initial Schmidt values are drawn from a truncated normal distribution, with an enforcement on the mean $\lambda_{mean}$, with standard deviations $\sigma = 0.01$ (top) and $\sigma = 0.07$ (bottom). The results are averaged over 10 different network samples. \hyperref[fig:fully_connected_honeycomb_b]{(d)}~Comparison of entanglement and integrity between quantum networks with initial equal Schmidt values (mean), and drawn from a truncated normal distribution with mean enforcement, with standard deviations $\sigma = 0.01, 0.07$.}
    \label{fig:fully_connected_honeycomb_plots}
\end{figure*}

Fig.~\hyperref[fig:diagonal_square_lattice_plot_b]{6c} shows the behavior of the average connectivity when percolating 10 quantum networks with random initial Schmidt values, plotted against the mean Schmidt value $\lambda_{mean}$. In the top plot, where Schmidt values are drawn from a truncated normal distribution with a standard deviation of $\sigma = 0.01$, the connectivity shifts seen in networks with uniform Schmidt values remain visible but appear noticeably smoothed out. In the bottom plot, where $\sigma = 0.07$, these shifts completely disappear, resulting in a fully smoothed connectivity curve. This phenomenon arises because larger quenched disorder among Schmidt values enables new local optimal solutions, unlocking alternative paths that would otherwise be suboptimal. This explanation is confirmed by Fig.~\hyperref[fig:diagonal_square_lattice_plot_b]{6d}, where we directly compare the average entanglement of the final state and the average integrity between quantum networks with equal initial Schmidt values and those with randomly drawn Schmidt values. While the entanglement curve remains similar in all three cases, integrity generally improves with increasing Schmidt value variance -- except for a small region around the first Schmidt value threshold. Notably, the transition point at which entanglement begins to decay remains unaffected, despite the presence of quenched disorder.

Fig.~\ref{fig:fully_connected_honeycomb_plots} shows the results for the fully connected honeycomb topology with 36 hexagons. The presence of different connection phases and the decay of entanglement after the last percolation threshold are also evident for this topology (Figs.~\hyperref[fig:fully_connected_honeycomb_a]{7a} and~\hyperref[fig:fully_connected_honeycomb_a]{7b}). Moreover, we observe the same dynamics when increasing the standard deviation between the initial Schmidt values of the quantum network (Figs.~\hyperref[fig:fully_connected_honeycomb_b]{7c} and~\hyperref[fig:fully_connected_honeycomb_b]{7d}). 

We highlight the presence of two additional connection phases compared to the diagonal square lattice, enabled by the higher connectivity of this topology. Knowing this, we can use the analytical description again to describe the strategies associated to these new phases, together with their corresponding percolation thresholds. The new strategies are the following:
\begin{itemize}
    \item Four swapping operations + distillation of the four resulting states. This strategy yields the threshold $\lambda_{th}^{4S,4D}~\approx~0.759$, solution of the inequality: 
    $$\left(\lambda_{SWAP}(\lambda, \lambda)\right)^4 \leq \frac{1}{2}$$
    \item Three swapping operations + distillation of the resulting states with a previously existing state. This strategy yields the threshold $\lambda_{th}^{3S, 4D}~\approx~0.779$, solution of the inequality:
    $$\left(\lambda_{SWAP}(\lambda, \lambda)\right)^3\cdot\lambda \leq \frac{1}{2}$$
\end{itemize}
Both the new strategies enabled by the fully connected honeycomb topology yield percolation thresholds larger than the last threshold for the diagonal square lattice. As a consequence, this topology is more robust than the former, as the entanglement decay starts at larger mean Schmidt value. However, the fully connected honeycomb topology is a much harder topology to set up, as its initial configuration requires many more connections.

\section{Conclusions}
\label{sec:conclusion}

We have considered a model of quantum network defined on a graph, where pure but non-maximally entangled states are initially shared between neighbor nodes~\cite{Acin2007EntPercolation}. Furthermore, we have introduced fluctuations in the entanglement shared between neighbors, as quantified through the Schmidt values. With the goal of establishing perfect entanglement between a pair of distant nodes, we have studied LOCC strategies that other nodes may implement to localize the entanglement in the given pair. In a regular lattice defined by an elementary cell, we have identified a hierarchy of local LOCC strategies that involve a different number of neighboring elementary cells. Given the modular structure of the considered networks, achieving perfect entanglement at finite distance also implies a strategy for entanglement percolation at longer range.

To test our analytical description and study quantum networks with large quenched disorder in their initial Schmidt values, we have developed a numerical framework that heuristically finds a path to percolate a quantum network to connect any pair of distant nodes. We analyzed the behavior of the entanglement of the final state and the connectivity of the network at the end of the percolation process for different initial parameters. We were able to distinguish various connection phases, which depend on the initial configuration of the network. Our framework enables an evaluation of the quality of a general quantum network with planar topology based on its resilience to entanglement disorder. Additionally, our analytical and numerical results yield new lower bounds on the entanglement percolation threshold, as expressed in terms of the mean Schmidt value of the entangled pairs initially shared by nearest neighbor sites.

This approach can be extended to random and non-regular networks~\cite{Brosco}, and may be applied to find new protocols for entanglement routing~\cite{nath2025generalconcurrencepercolationquantum}, or in error detection in distributed quantum computing applications~\cite{pattison2024fastquantuminterconnectsconstantrate}. Furthermore, our methodology can be extended to incorporate dynamical effects and constraints in the use of local resources~\cite{PhysRevLett.134.030803}, or to simulate quantum networks in higher dimensions~\cite{PhysRevA.81.012310}. Finally, it would be insightful to apply our approach to more physically accurate models of quantum networks, where links between node pairs are represented by mixed states~\cite{lapeyre2012distributionentanglementnetworksbipartite, Broadfoot_2009}. In particular, we could compare our numerical results for pure-state quantum networks with those obtained using methodologies specifically designed for mixed-state quantum networks~\cite{bala2025statisticalanalysismultipathentanglement}. Notably, existing results from mixed-state models indicate that percolation strategies can still be detected, suggesting that our analytical description may extend naturally to these more general scenarios.

\section*{Acknowledgments}
\label{sec:acknowledgment}

We thank Zixin Huang for many useful comments.
This work is supported by: 
the European Union's Horizon Europe research and innovation programme under the Project ``Quantum Secure Networks Partnership'' (QSNP, Grant Agreement No.~101114043); 
the European Union's NextGenerationEU
through 
projects 
``CN00000013 - Italian Research Center on HPC, Big Data and Quantum Computing (ICSC)'' and ``PE0000023  - National Quantum Science and Technology Institute (NQSTI)'';
PRIN2022 ``2022NZP4T3 - QUEXO'' (CUP D53D23002850006);
the Italian Ministry of University and Research (MUR)
via the Department of Excellence grants 2023-2027 project ``Quantum Sensing and Modelling for One-Health (QuaSiModO)''; the Italian ``Istituto Nazionale di Fisica Nucleare (INFN)'', via the initiative IS-QUANTUM; the University of Bari, via the 2023-UNBACLE-0244025 grant. A.D.G. acknowledges financial support from the European Union's Horizon Europe research and innovation programme (Quantum Flagship) under the projects PASQuanS2 (Grant Agreement No.~101113690) and EuRyQa (Grant Agreement No.~101070144). The authors acknowledge computational resources by CINECA and the University of Bari and INFN cluster RECAS~\cite{ReCaS}.

\appendix
\section{Majorization theory and its connection to entanglement distillation}
\label{app:majorization}

Consider two vectors $\vec{v}^\downarrow, \vec{w}^\downarrow$, each with $d$ positive entries sorted in descending order and satisfying $\displaystyle \sum_i v^\downarrow_i =\sum_i w^\downarrow_i = 1$. Then, vector $\vec{v}^\downarrow$ is said to be \textit{majorized} by $\vec{w}^\downarrow$, or $\vec{v}^\downarrow < \vec{w}^\downarrow$, if and only if:
\begin{equation}
    \displaystyle \sum_{i=1}^l v_i^\downarrow \leq \sum_{i=1}^l w_i^\downarrow, \forall \ l \in [1, d]
\end{equation}
Furthermore, $\vec{v}^\downarrow$ is said to be \textit{submajorized} by $\vec{v}^\downarrow$, or $\vec{v}^\downarrow <^w \vec{w}^\downarrow$, if and only if:
\begin{equation}
    \displaystyle \sum_{i=l}^d v_i^\downarrow \geq \sum_{i=l}^d w_i^\downarrow, \forall \ l \in [1, d]
\end{equation}

\begin{figure*}[t!]
    \centering
    \includegraphics[width=0.9\linewidth]{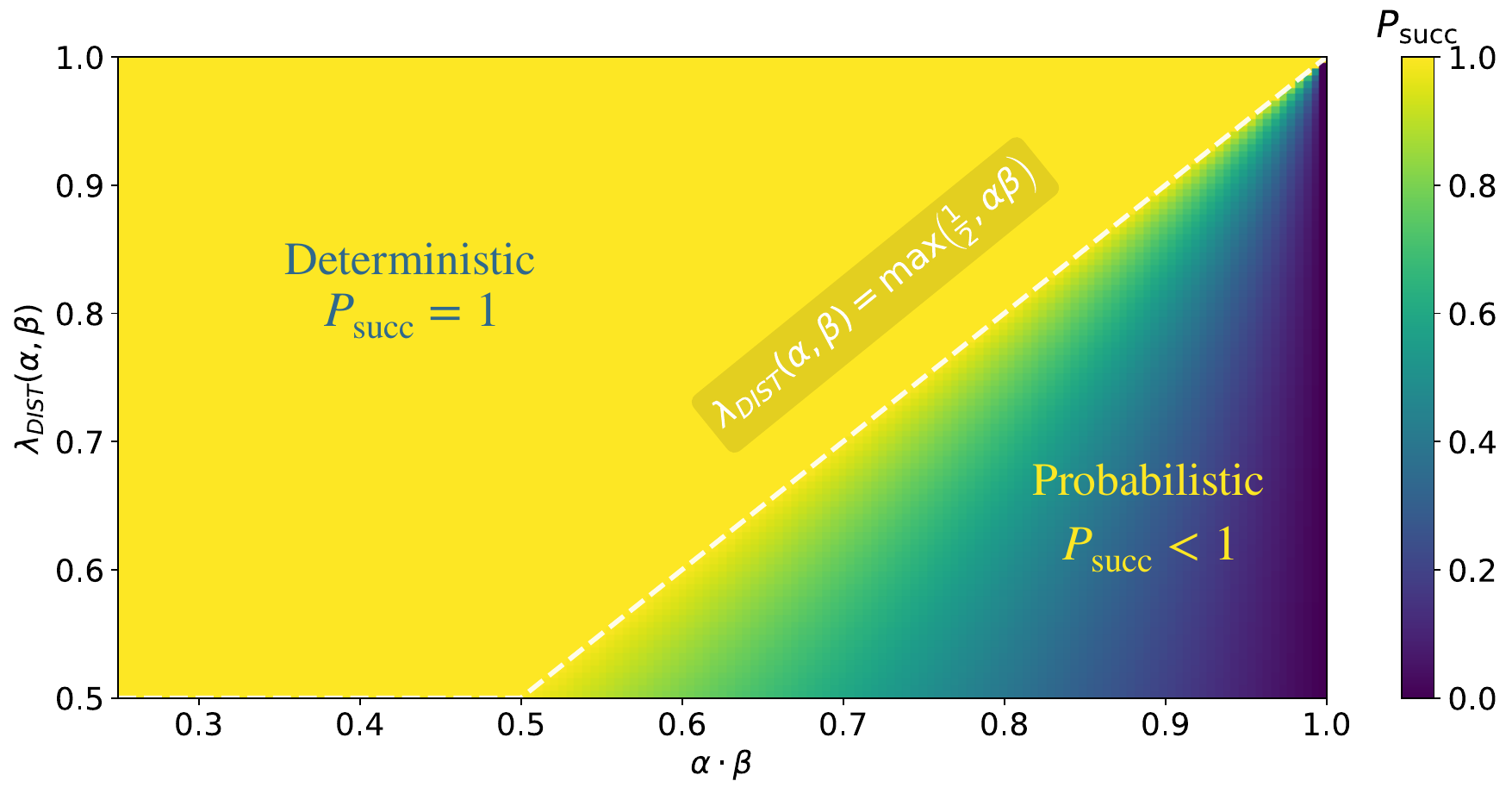}
    \caption{Success probability $P_\mathrm{succ} = \min\left(1, \frac{1-\alpha\beta}{1-\lambda_{DIST}(\alpha, \beta)}\right)$ for converting a state $\ket{\alpha}\otimes\ket{\beta}$ into a state $\ket{\lambda_{DIST}(\alpha,\beta)}\otimes\ket{00}$, plotted as a function of the product of the initial Schmidt values $\alpha\cdot\beta$ and the largest Schmidt value of the target state $\lambda_{DIST}(\alpha,\beta)$. The white contour line corresponds to the threshold $\lambda_{DIST}(\alpha, \beta) = \max\left(\frac{1}{2}, \alpha\beta\right)$, which separates the regions where $P_{\mathrm{succ}} = 1$ (above the line) and $P_{\mathrm{succ}} < 1$ (below the line).}
    \label{fig:p_succ}
\end{figure*}

Nielsen was the first to notice the connection between majorization theory and quantum information~\cite{Nielsen_1999}. Consider the problem of transforming a general pure state $\ket{\psi}$ into another state $\ket{\phi}$ using LOCC. Define $\vec{\psi}^\downarrow$ and $\vec{\phi}^\downarrow$ as the vectors of Schmidt values of states $\ket{\psi}$ and $\ket{\phi}$ respectively. Then, state $\ket{\psi}$ can be deterministically transformed into $\ket{\phi}$ via LOCC if and only if:
\begin{equation}
    \vec{\psi}^\downarrow < \vec{\phi}^\downarrow
    \label{eq:NielsenTh}
\end{equation}
A generalization of this result was proven by Vidal in~\cite{PhysRevLett.83.1046}, and it states that any state $\ket{\psi}$ can be converted into another state $\ket{\phi}$ via LOCC with a certain success probability $P_{\mathrm{succ}}$ if and only if:
\begin{equation}
    \vec{\psi}^\downarrow <^\phi P_{\mathrm{succ}}\vec{\phi}^\downarrow
    \label{eq:VidalTh}
\end{equation}
The above equation can be restated as follows: the best probability of success $P_{\mathrm{succ}}$ to convert a state $\ket{\psi}$ into another state $\ket{\phi}$ is
\begin{equation}
    P_\mathrm{succ} = \displaystyle \min_{l\in [1,d]} \left(\frac{\sum_{i=l}^d\psi_l^\downarrow}{\sum_{i=l}^d\phi_l^\downarrow}\right)
    \label{eq:VidalThP}
\end{equation}

We will now use Eq.~\ref{eq:VidalThP} to prove that the combined states $\ket{\alpha}\otimes\ket{\beta}$ can be converted into $\ket{\lambda_{DIST}(\alpha, \beta)}$ deterministically. Moreover, we will show that $\ket{\lambda_{DIST}(\alpha, \beta)}$ defined in Eq.~\ref{eq:state_dist} is the state with the largest amount of entanglement that can be obtained deterministically. Define $\psi^\downarrow$ as the vector of Schmidt values, in descending order, of the state
\begin{align}
    \ket{\alpha}\otimes\ket{\beta} = (\sqrt{\alpha}\ket{00} + &\sqrt{1-\alpha}\ket{11}) \\ &\otimes(\sqrt{\beta}\ket{00}+\sqrt{1-\beta}\ket{11})\nonumber
\end{align}
Analogously, define $\phi^\downarrow$ as the set of Schmidt values, in descending order, of the target state
\begin{align}
    \ket{\alpha}\otimes\ket{\beta} &= \Big(\sqrt{\lambda_{DIST}(\alpha, \beta)} \ket{00} + \nonumber \\
    &\quad \sqrt{1-\lambda_{DIST}(\alpha, \beta)} \ket{11}\Big) \nonumber \\
    &\quad \otimes\left(\sqrt{\tilde{\phi}}\ket{00}+\sqrt{1-\tilde{\phi}}\ket{11}\right)\nonumber
\end{align}
with $\lambda_{DIST}(\alpha, \beta)$ the largest Schmidt value of the new entangled state. As our objective is to concentrate the entanglement of the original states into a single, more entangled state, we will be working in the limit of $\tilde{\phi} \rightarrow 1$. The vector $\psi^\downarrow$ can be written as
\begin{align}
    \psi^\downarrow = (&\alpha\beta, \max(\alpha, \beta)(1-\min(\alpha, \beta)), \nonumber\\ &\min(\alpha, \beta)(1-\max(\alpha, \beta)), (1-\alpha)(1-\beta))
\end{align}
Equivalently for $\phi^\downarrow$,
\begin{align}
    \phi^\downarrow = (&\lambda_{DIST}(\alpha, \beta)\tilde{\phi}, (1-\lambda_{DIST}(\alpha, \beta))\tilde{\phi}, \nonumber \\ &\lambda_{DIST}(\alpha, \beta)(1-\tilde{\phi}), (1-\lambda_{DIST}(\alpha, \beta))(1-\tilde{\phi}))
\end{align}
Let us then use Vidal's theorem in the limit $\tilde{\phi}\rightarrow 1$:
\begin{align}
    P_{\mathrm{succ}} & = \displaystyle \lim_{\tilde{\phi}\rightarrow 1} \min\Biggl(1, \frac{1-\alpha\beta}{(1-\lambda_{DIST}(\alpha, \beta))\tilde{\phi}}, \frac{1-\max(\alpha, \beta)}{1-\tilde{\phi}}, \nonumber \\
    & \hspace{3.5cm} \frac{(1-\alpha)(1-\beta)}{(1-\lambda_{DIST}(\alpha, \beta))(1-\tilde{\phi})}\Biggl) \nonumber \\
    & = \min\left(1, \frac{1-\alpha\beta}{1-\lambda_{DIST}(\alpha, \beta)}\right)
    \label{eq:psucc}
\end{align}
Let us now choose $\lambda_{DIST}(\alpha, \beta) = \max\left(\frac{1}{2}, \alpha\beta\right)$ as in Eq.~\ref{eq:lambda_dist} and split our analysis for each of the two cases. If $\alpha\beta > \frac{1}{2}$, $\lambda_{DIST}(\alpha, \beta) = \alpha\beta$, meaning the success probability becomes
\begin{equation}
    P_{\mathrm{succ}} = \min\left(1, \frac{1-\alpha\beta}{1-\alpha\beta}\right) = 1
\end{equation}
If instead $\alpha\beta \leq \frac{1}{2}$, $\lambda_{DIST}(\alpha, \beta) = \frac{1}{2}$, meaning the success probability can be computed as
\begin{equation}
    P_{\mathrm{succ}} = \min\left(1, 2(1-\alpha\beta)\right)
\end{equation}
Since $\alpha\beta \leq \frac{1}{2}$, it immediately follows $2(1-\alpha\beta) \geq 1 \implies P_{\mathrm{succ}} = 1$, which concludes our proof. 

As a final remark, we analyze how the success probability $P_\mathrm{succ}$ computed in Eq.~\ref{eq:psucc} varies with different values of $\lambda_{DIST}(\alpha, \beta)$ and the product $\alpha\cdot\beta$. Fig.~\ref{fig:p_succ} highlights the separation between a deterministic regime, where $P_{\mathrm{succ}} = 1$, and a probabilistic regime, where generating states with a certain amount of entanglement is no longer guaranteed. The separating contour corresponds to the line $\lambda_{DIST}(\alpha, \beta) = \max\left(\frac{1}{2}, \alpha\beta\right)$, which indicates the most entangled state that can still be prepared deterministically.

\section{Example application of physics-informed heuristics}
\label{app:example_2}
In this Appendix, we expand on the example from Fig.~\ref{fig:percol_strat_example}, completing the local-strategy-finding routine and explaining the employed heuristics on a practical example.

Fig.~\ref{fig:percol_strat_example_2} shows the successive steps of the path-finding algorithm. At the end of the first local-strategy-finding routine, the algorithm selects node $N_2$ as the optimal local solution. However, this choice quickly proves to be globally suboptimal, as no more local percolation strategies involving distillation are available along the remaining path to the target node. As a consequence, the final state between the source and target nodes is not maximally entangled (Fig.~\hyperref[fig:psexample2_c]{9c}). To improve the overall entanglement, the algorithm then iterates over all non-maximally entangled states generated during the process, searching for alternative paths that could increase entanglement through additional distillation steps. In this case, the state connecting $N_5$ to $T$ can be improved by performing two swaps on $N_3$ and $N_4$, followed by a distillation step with the existing state on the path. While this procedure successfully yields a maximally entangled state between $N_5$ and $T$, it only slightly improves the entanglement of the final state between $S$ and $T$, as the state between $N_2$ and $N_5$ cannot be further distilled. This shows that the locally optimal choice of $N_2$ actually leads to a dead end, disallowing a perfect connection between the source and target nodes.

In the next sample, we vary our heuristic parameters to consider not only optimal local solutions, but also slightly suboptimal ones. Suppose the algorithm now selects $N_3$ as the next starting node. In this case, $N_3$ is connected with a maximally entangled state to the source via swapping $N_1$ and $N_2$ and distilling the resulting three states (see Fig.~\hyperref[fig:psexample_c]{4c}). This solution was discarded in the previous sample, as it destroys more states than the path to $N_2$. However, we now find that not only is a direct connection between $N_3$ and $T$ possible via a local percolation strategy, but the two available swapping paths also enable the generation of a maximally entangled state between them (Fig.~\hyperref[fig:psexample2_e]{9e}). Finally, by performing a swap on $N_3$, the algorithm is able to connect nodes $S$ and $T$ with a maximally entangled state (Fig.~\hyperref[fig:psexample2_f]{9f}). This example demonstrates that deviating from the locally optimal solution can, in some cases, lead to the globally optimal outcome.

\begin{figure*}[ht!]
  \begin{subfigure}{0.3\textwidth}
      \caption{}
      \label{fig:psexample2_a}
      \centering
      \includegraphics[width=5.5cm]{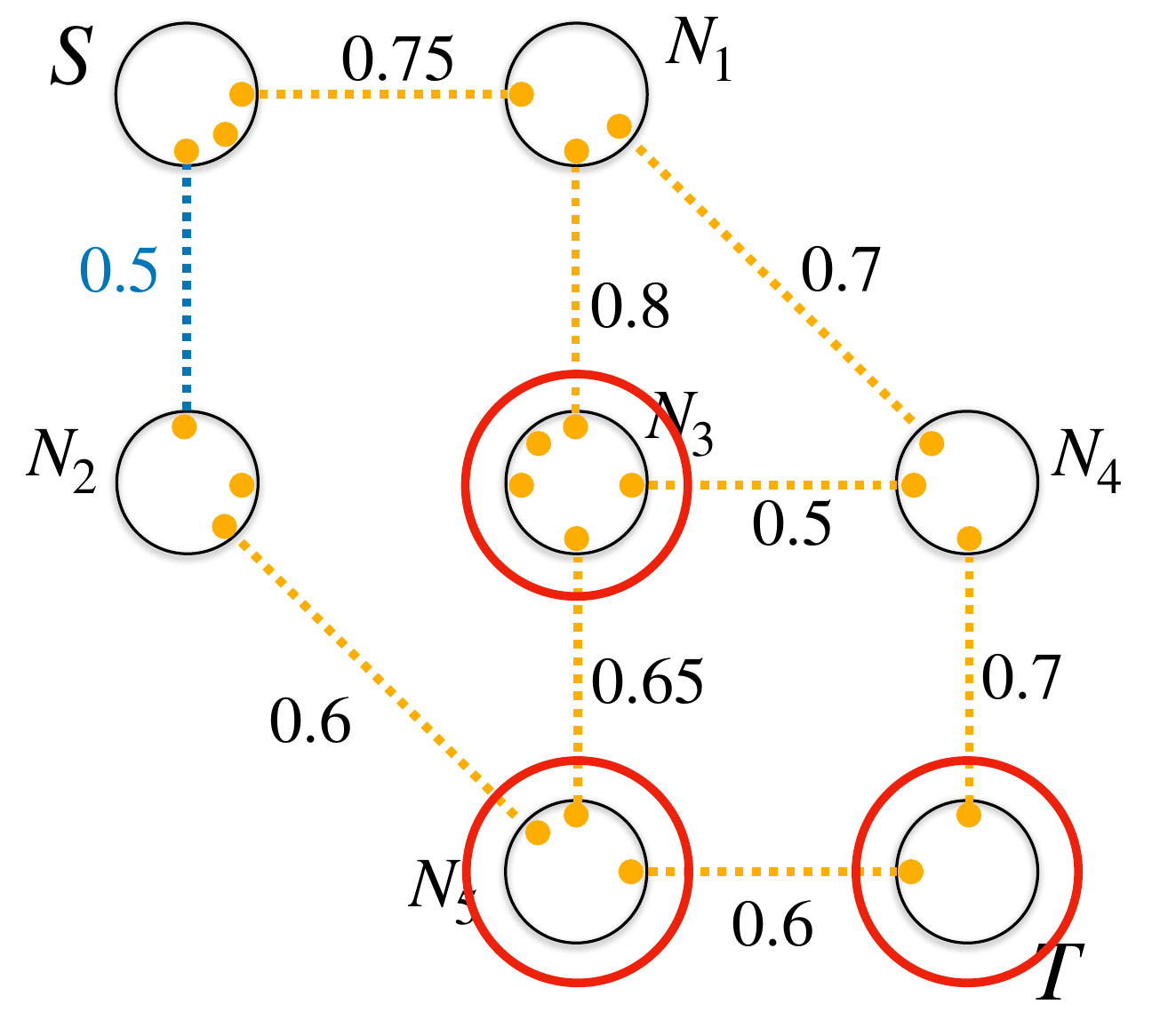}
  \end{subfigure}
  \hspace{1mm}
  \begin{subfigure}{0.3\textwidth}
      \caption{}
      \label{fig:psexample2_b}
      \centering
      \includegraphics[width=5.5cm]{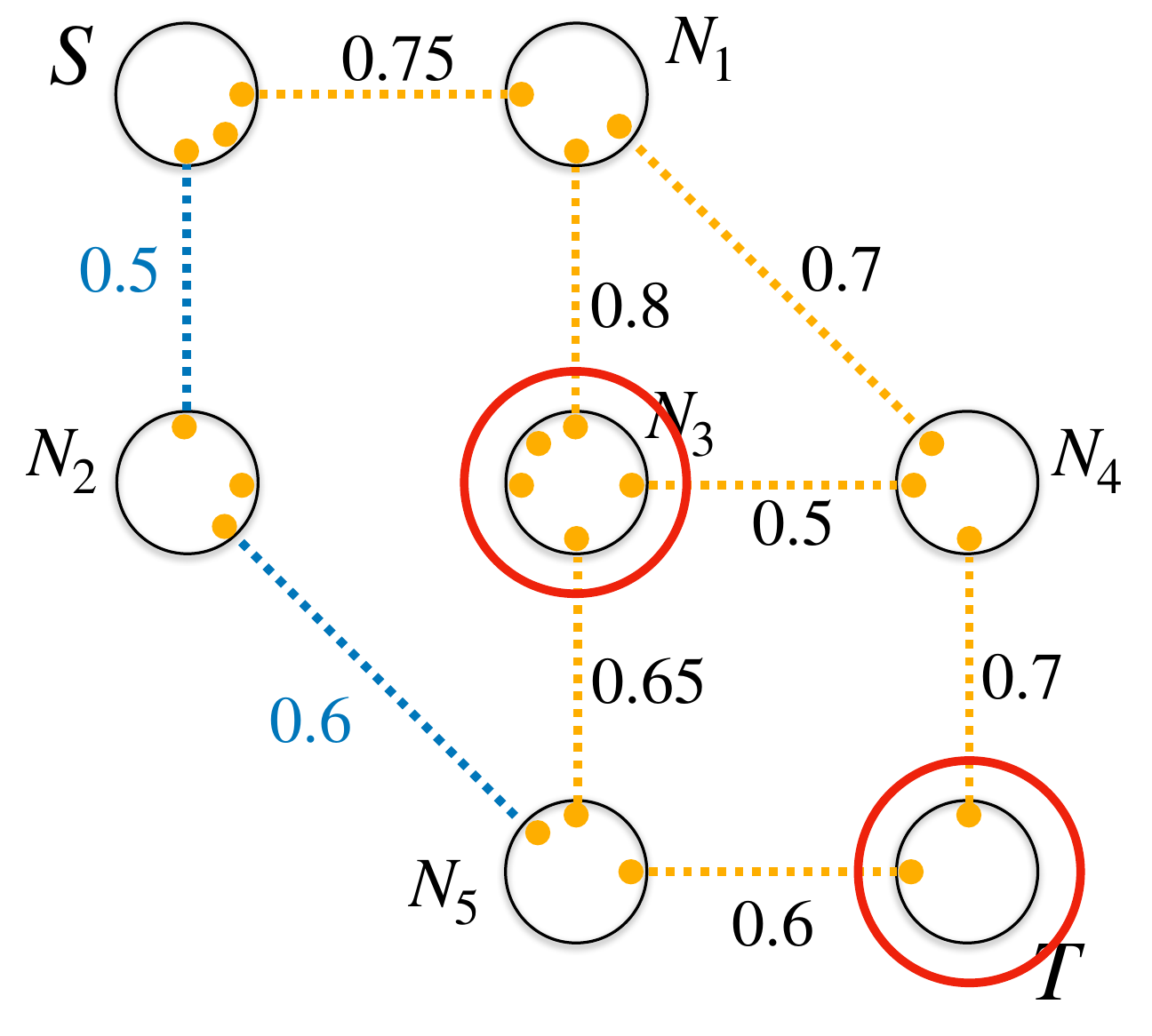}
  \end{subfigure}
  \hspace{1mm}
  \begin{subfigure}{0.3\textwidth}
      \caption{}
      \label{fig:psexample2_c}
      \centering
      \includegraphics[width=5.5cm]{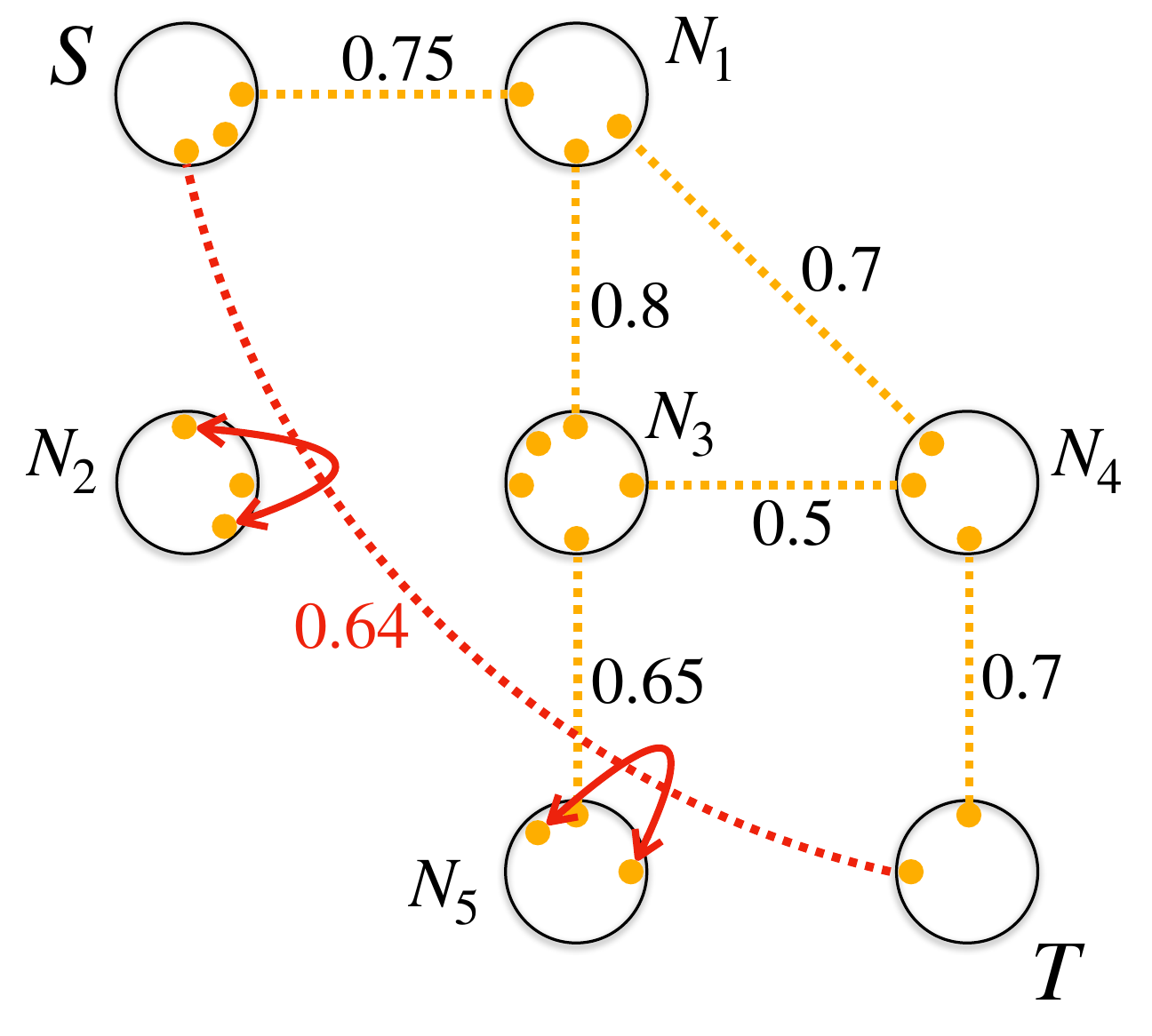}
  \end{subfigure}
  \hspace{1mm}
  \begin{subfigure}{0.3\textwidth}
      \caption{}
      \label{fig:psexample2_d}
      \centering
      \includegraphics[width=5.5cm]{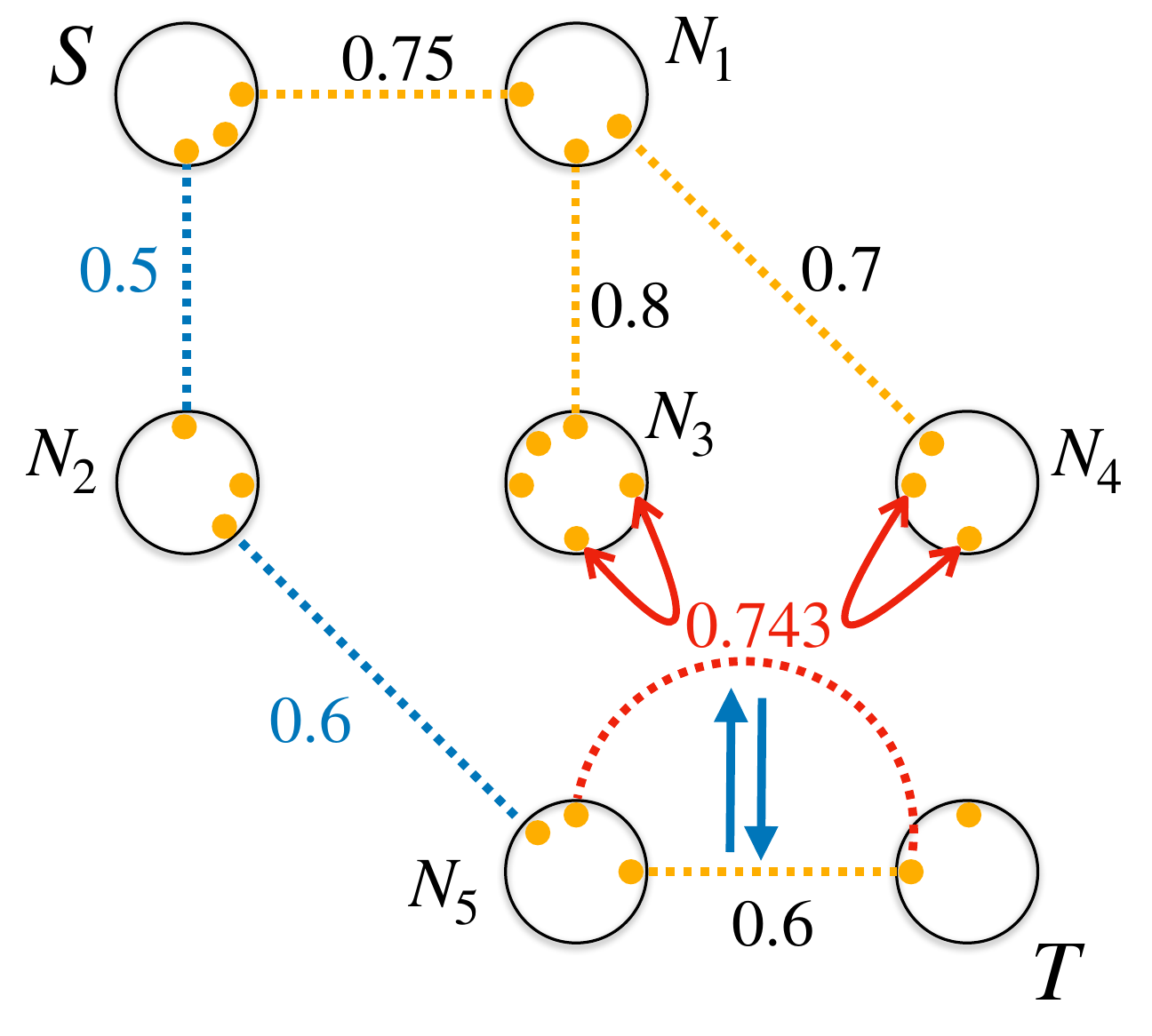}
  \end{subfigure}
  \hspace{1mm}
  \begin{subfigure}{0.3\textwidth}
      \caption{}
      \label{fig:psexample2_e}
      \centering
      \includegraphics[width=5.5cm]{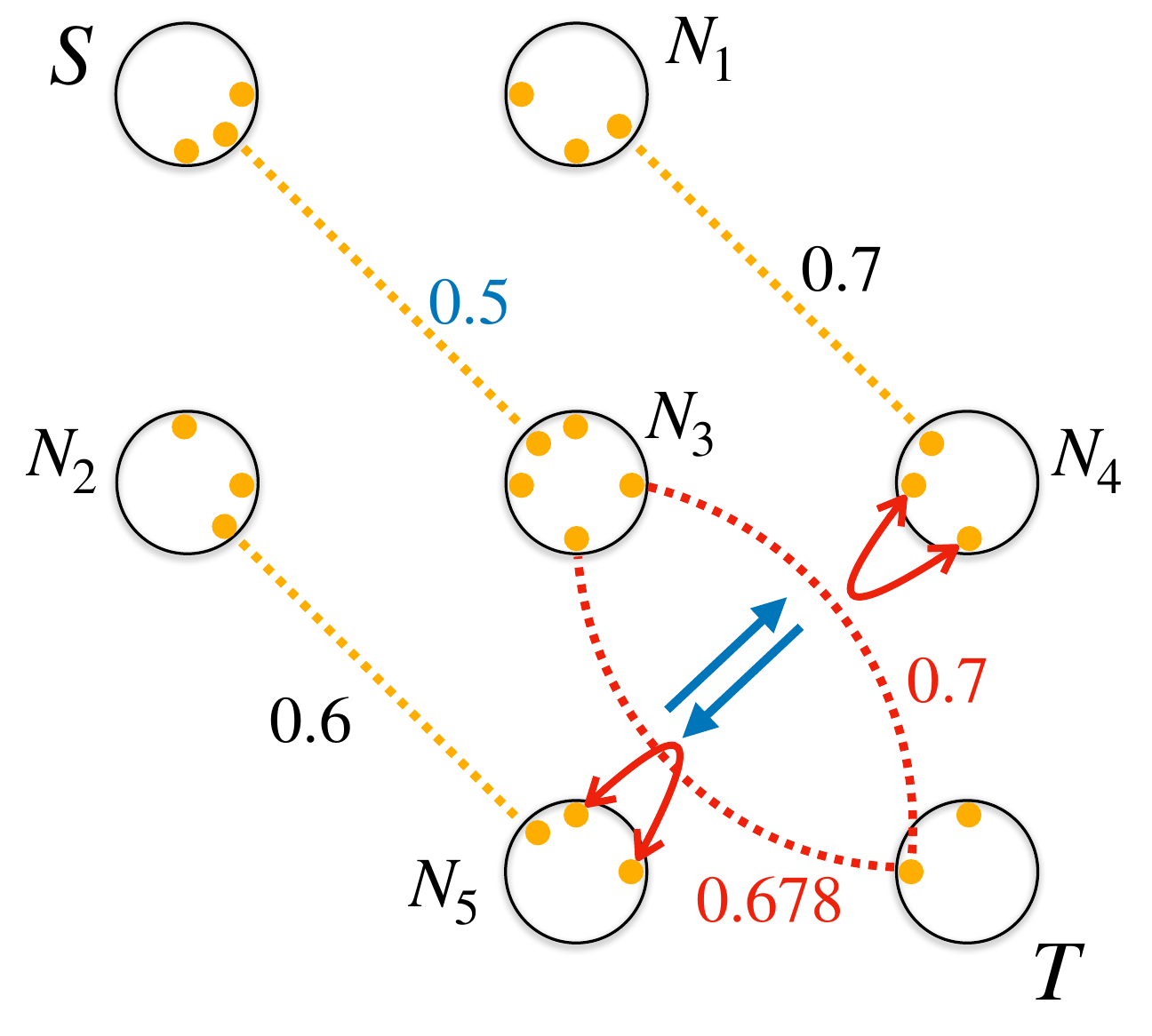}
  \end{subfigure}
  \hspace{1mm}
  \begin{subfigure}{0.3\textwidth}
      \caption{}
      \label{fig:psexample2_f}
      \centering
      \includegraphics[width=5.5cm]{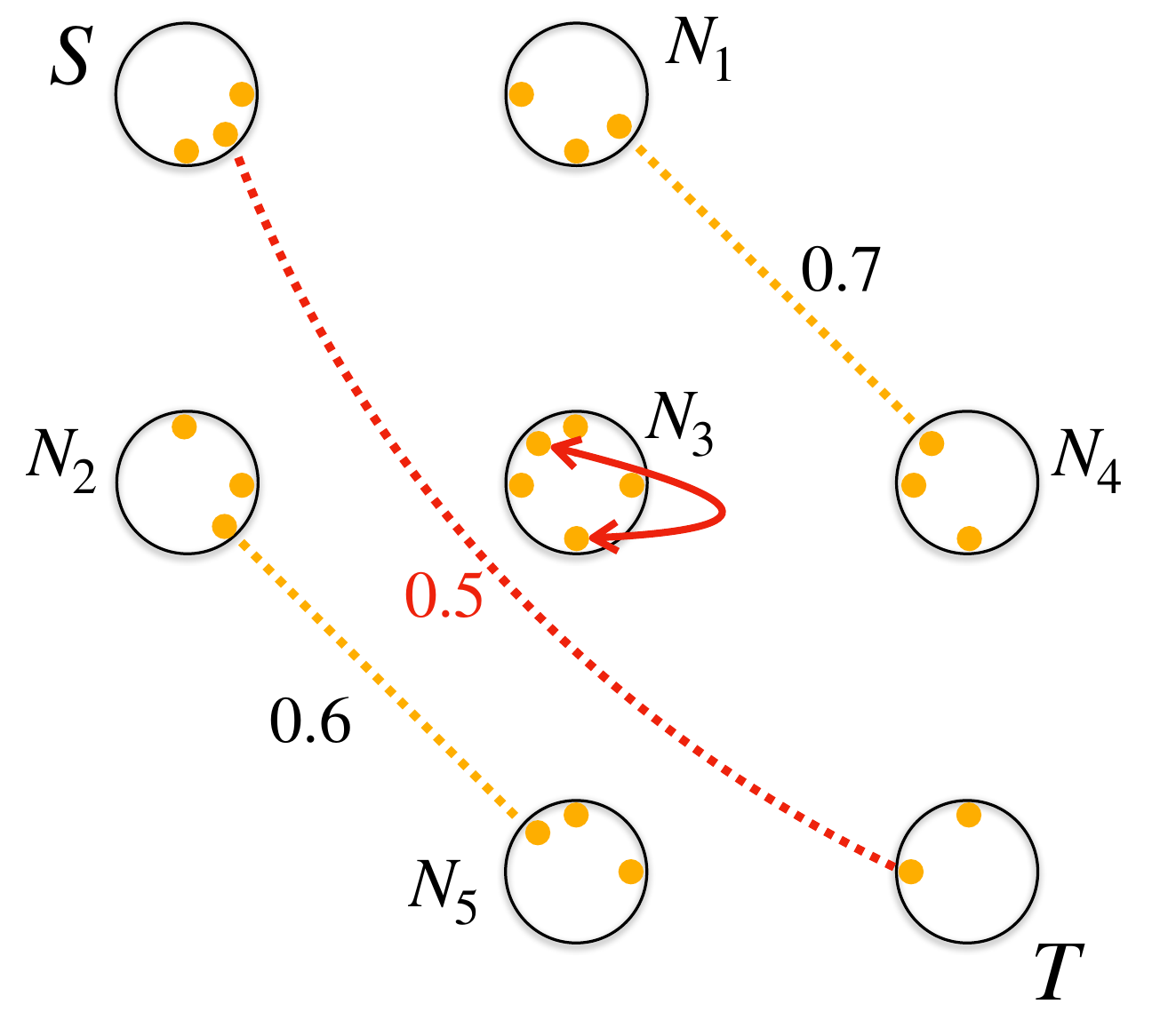}
  \end{subfigure}
  \caption{Continued schematic of the iteration process of the local-strategy-finding routine and heuristic exploration of alternative paths for a quantum network with 7 nodes. \hyperref[fig:psexample_a]{(a)}~Having chosen $N_2$ as the local optimal solution, the algorithm analyzes all nodes at distance at most $2$ from $N_2$. There are three nodes available, but none of them can be reached via combined swapping or distillation. The node that maximizes the entanglement is $N_5$, which is chosen as the optimal solution. \hyperref[fig:psexample_b]{(b)}~From node $N_5$ we can reach the target node, but without any local distillation to improve their state. Nonetheless, this is the optimal local solution at this point. \hyperref[fig:psexample_c]{(c)}~Once the target node has been reached, we perform entanglement swapping on each of the nodes on the selected path. In this case, the final state is not maximally entangled, meaning the algorithm starts looking for alternative paths around the non-maximally entangled states selected throughout the procedure. \hyperref[fig:psexample_d]{(d)}~By iterating over alternative nodes surround $N_5$, the algorithm eventually finds an alternative path to connect nodes $N_5$ and the target node $T$, via $N_3$ and $N_4$, which creates a maximally entangled state between the two nodes. Unfortunately, this is still not enough to perfectly connect the source to the target, as the state connecting $N_2$ and $N_5$ is not maximally entangled. \hyperref[fig:psexample_e]{(e)}~In the next sample, the algorithm chooses the suboptimal local solution that connects $S$ to $N_3$. From $N_3$, it is possible to directly move to the target node by distilling two swapping paths through $N_4$ and $N_5$. \hyperref[fig:psexample_f]{(f)}~Finally, since we reached the target node, we perform swapping on $N_3$ to connect the source to the target node with a maximally entangled state.}
    \label{fig:percol_strat_example_2}
\end{figure*}

\section{Effect of random distribution of entanglement in a local quantum percolation strategy}
\label{app:deviations}

When all states of the network are assumed to yield the same Schmidt value $\lambda$, we can express the swapping and distillation operations in the following way:
\begin{equation}
    \epsilon_{SWAP}(\epsilon, \epsilon) = \sqrt{2\epsilon^2 - 4\epsilon^4}
    \label{eq:eps_swap_equal}
\end{equation}
\begin{equation}
    \epsilon_{DIST}(\epsilon, \epsilon) = \max\left\{0, \epsilon^2 + \epsilon - \frac{1}{4}\right\}
    \label{eq:eps_distill_equal}
\end{equation}
Moreover, Eq.~\ref{eq:char_eq} can be simplified as follows:
\begin{align}
    \displaystyle \prod_{\left(j_1, j_2\right) \in \mathbf{S}} \lambda_{SWAP}\left(\lambda, \lambda\right) \cdot \prod_{i \in \mathbf{K}} \lambda = \left(\lambda_{SWAP}\left(\lambda, \lambda\right)\right)^s\cdot\lambda^k
    \label{eq:char_eq_allequal}
\end{align}
where $s = \lvert \mathbf{S} \rvert$ and $k = \lvert \mathbf{K} \rvert$.
The solution of this inequality, i.e., the percolation threshold of the strategy, can be easily computed in this case. However, in the presence of quenched entanglement disorder in the network, its initial states do not yield the same Schmidt values. In this setup, it is not trivial to understand how the percolation thresholds are affected. Therefore, we study pure-state quantum networks with different Schmidt values by identifying the mean Schmidt value $\lambda_{mean}$ of the states employed in a local percolation strategy and evaluating the deviation of each Schmidt value from this mean. Using this convention, we can define each value $\lambda_i \in \mathbf{S} \cup \mathbf{K}$ as $\lambda_{i} := \lambda_{mean} + \lambda_{\sigma, i}$, with $\lambda_{mean} := \dfrac{1}{s+k}\displaystyle\sum_{i \in \mathbf{S} \cup \mathbf{K}}\lambda_i$, $\lambda_{mean} \in [0, \frac{1}{2}]$, and $\lambda_{\sigma_i}~\in~(-\frac{1}{2}, \frac{1}{2})$. Equivalently, $\lambda_i = \frac{1}{2} + \epsilon_i$ and $\epsilon_i = \epsilon_{mean} + \epsilon_{\sigma, i}$.

First, we try to understand the effect of randomly distributed entanglement on swapping and distillation operations. For this purpose, we can rewrite Eq.~\ref{eq:lambda_swap_eps} with the above definition of $\epsilon_i$, and using $\epsilon_{mean}^{i,j}~:=~\frac{\epsilon_i + \epsilon_j}{2}$ and $\epsilon_\sigma~:=~\epsilon_{\sigma,i}~=~-~\epsilon_{\sigma,j}$:
\begin{align}
    &\epsilon_{SWAP}\left(\epsilon_i, \epsilon_j\right) = \nonumber \\ &= \epsilon_{SWAP}\left(\epsilon_{mean} + \epsilon_{\sigma,i}, \epsilon_{mean}^{i,j} + \epsilon_{\sigma,j}\right) \nonumber \\ &= \epsilon_{SWAP}\left(\epsilon_{mean}^{i,j} + \lvert \epsilon_\sigma \rvert, \epsilon_{mean}^{i,j} - \lvert \epsilon_\sigma \rvert\right) \nonumber \\ &= \sqrt{\underbrace{2\left(\epsilon_{mean}^{i,j}\right)^2 - 4\left(\epsilon_{mean}^{i,j}\right)^4}_{=:\left(\epsilon^{SWAP}_{mean}\right)^2}+ \underbrace{2\epsilon_{\sigma}^2 - 4\epsilon_{\sigma}^4 + 8 \left(\epsilon_{mean}^{i,j}\right)^2\epsilon_{\sigma}^2}_{=: \left(\epsilon_{\sigma}^{SWAP}\right)^2\geq0}}
\end{align}
Comparing this expression with the case of states with equal entanglement (Eq.~\ref{eq:eps_swap_equal}), we notice an additional positive term inside the square root, meaning $\epsilon_{SWAP}$ increases with larger $\epsilon_{\sigma}$. Consequently, when the initial states $i$ and $j$ are not maximally entangled, the swapping operation yields a less entangled state the greater the distance between the Schmidt values $\lambda_i$ and $\lambda_j$. 

Let us repeat the same procedure for distillation:
\begin{align}
    \epsilon_{DIST}&\left(\epsilon_{mean}^{i,j} + \lvert \epsilon_\sigma \rvert, \epsilon_{mean}^{i,j} - \lvert \epsilon_\sigma \rvert\right) = \nonumber \\ &= \left(\epsilon_{mean}^{i,j}\right)^2 + \epsilon_{mean}^{i,j} - \frac{1}{4} - \epsilon_{\sigma}^2
\end{align}
Comparing again to Eq.~\ref{eq:eps_distill_equal}, we notice that the additional term depending on $\epsilon_\sigma$ is always negative, meaning that $\epsilon_{DIST}$ decreases with larger $\epsilon_\sigma$. It follows that distillation improves the entanglement of the state if the deviation is higher. This fact is confirmed by the AM-GM inequality, which follows:
\begin{equation}
    \sqrt[n]{\displaystyle \prod_{i=1}^n \lambda_i} \leq \overbrace{\frac{1}{n}\sum_{i=1}^n \lambda_i}^{=:\lambda_{mean}} \implies \displaystyle \prod_{i=1}^n \lambda_i \leq \lambda_{mean}^n
\end{equation}
The expression above means that the improving effect of deviations in distillations is generalized for the distillation of multiple states. Moreover, we also know that the inequality becomes an equality only when all states in the network are initially equal. As a consequence, the general inequality (Eq.~\ref{eq:char_eq}) is actually upper bounded by the expression on the left side of Eq.~\ref{eq:char_eq_allequal}:
\begin{align}
    \displaystyle \prod_{\left(j_1, j_2\right) \in \mathbf{S}} &\lambda_{SWAP}\left(\lambda_{j_1}, \lambda_{j_2}\right) \cdot \prod_{i \in \mathbf{K}} \lambda_{i} \nonumber \\ &\leq \left(\lambda_{SWAP}\left(\lambda_{mean}, \lambda_{mean}\right)\right)^s\cdot\lambda_{mean}^k
\end{align}

\section{Results for square and honeycomb lattices}
\label{app:square_honeycomb}

\begin{figure*}[ht!]
  \begin{subfigure}{0.45\textwidth}
      \label{fig:square_lattice_plot_a}
      \centering
      \includegraphics[width=8.2cm]{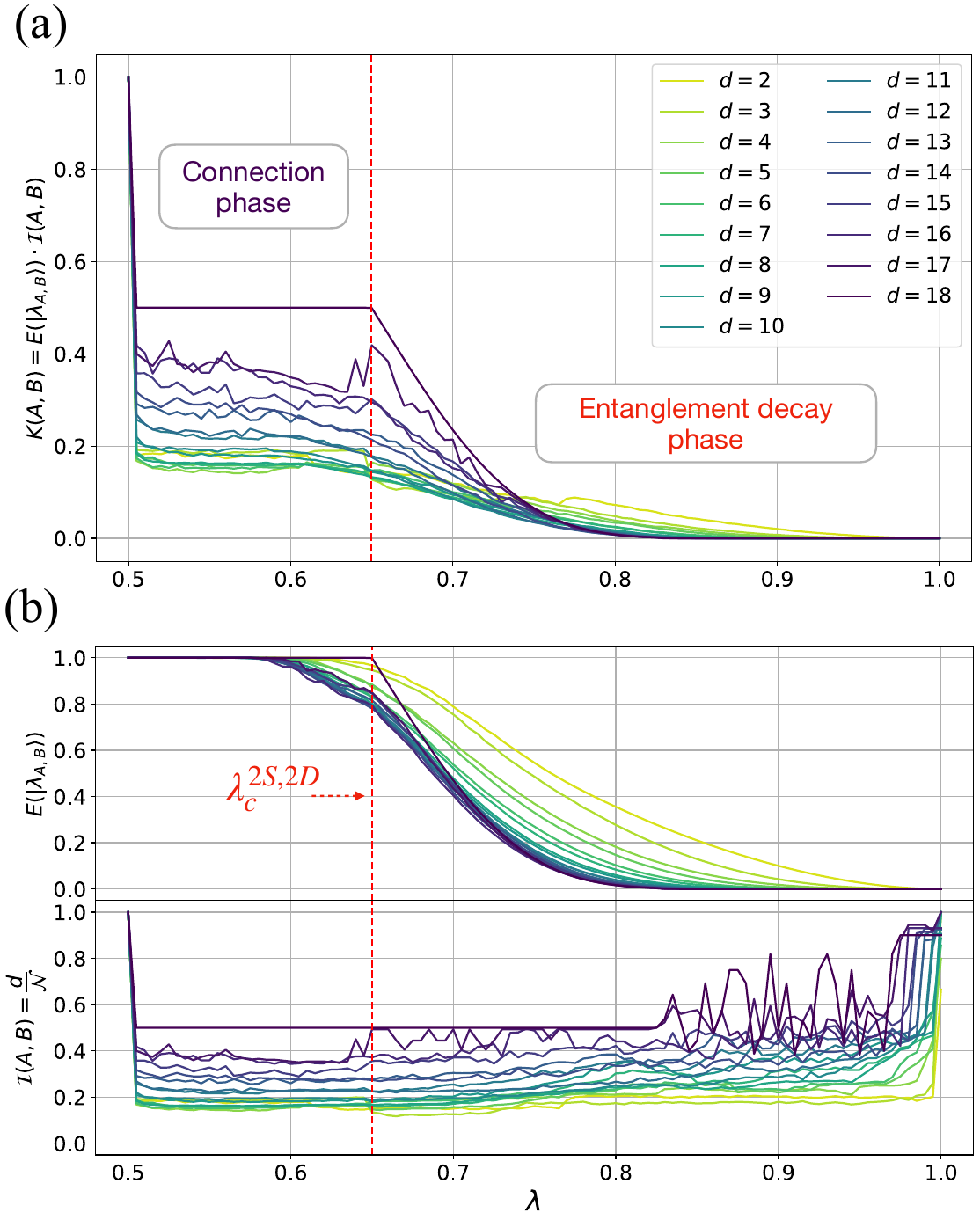}
  \end{subfigure}
  \hspace{5mm}
  \begin{subfigure}{0.45\textwidth}
      \label{fig:square_lattice_plot_b}
      \centering
      \includegraphics[width=8.2cm]{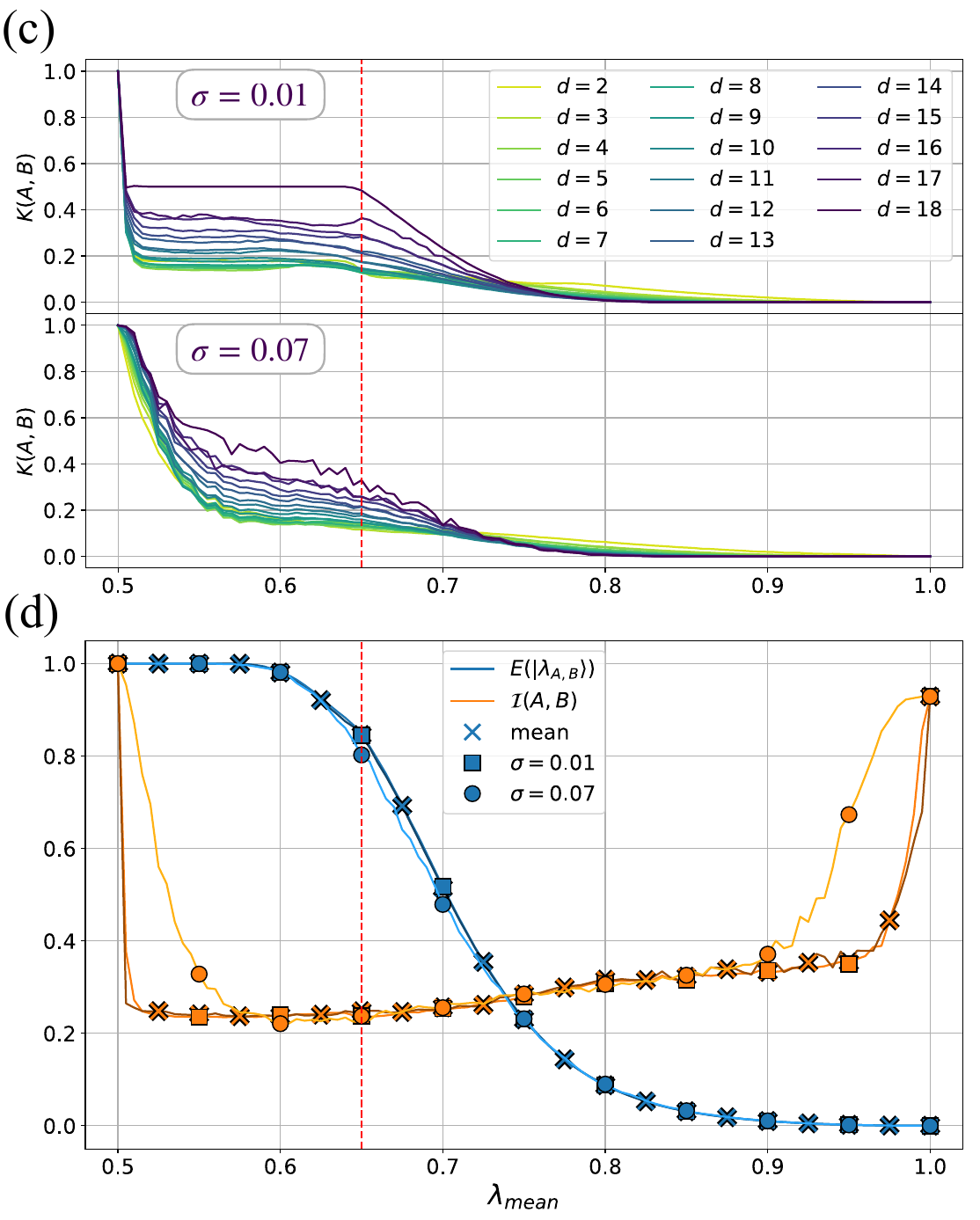}
  \end{subfigure}
  \caption{Results for a 10x10 square lattice quantum network. \hyperref[fig:square_lattice_plot_a]{(a)}~Connectivity of the network after quantum percolation between all possible distant node pairs, plotted against the Schmidt value assigned to all the initial states of the network. Each curve corresponds to the average connectivity of all pairs of nodes at a distance $d$. The vertical dashed line indicates the percolation threshold $\lambda_{th}^{2S,2D}$, predicted with the analytical description. \hyperref[fig:square_lattice_plot_a]{(b)}~Entanglement of the final state between distant nodes $A$ and $B$ (above) and integrity of the network (below) after quantum percolation between all possible pairs at distance $d$, plotted against the Schmidt value assigned to all the initial states of the network. \hyperref[fig:square_lattice_plot_b]{(c)}~Connectivity of the network after quantum percolation between all possible pairs at distance $d$, plotted against the mean Schmidt value of the initial states of the network. The initial Schmidt values are drawn from a truncated normal distribution, with an enforcement on the mean $\lambda_{mean}$, with standard deviations $\sigma = 0.01$ (above) and $\sigma = 0.07$ (below). The results are averaged over 10 different network samples. \hyperref[fig:square_lattice_plot_b]{(d)}~Comparison of entanglement and integrity between quantum networks with initial equal Schmidt values (mean), and drawn from a truncated normal distribution with mean enforcement, with standard deviations $\sigma = 0.01, 0.07$.}
    \label{fig:square_lattice_plots}
\end{figure*}

\begin{figure*}[ht!]
  \begin{subfigure}{0.45\textwidth}
      \label{fig:honeycomb_a}
      \centering
      \includegraphics[width=8.2cm]{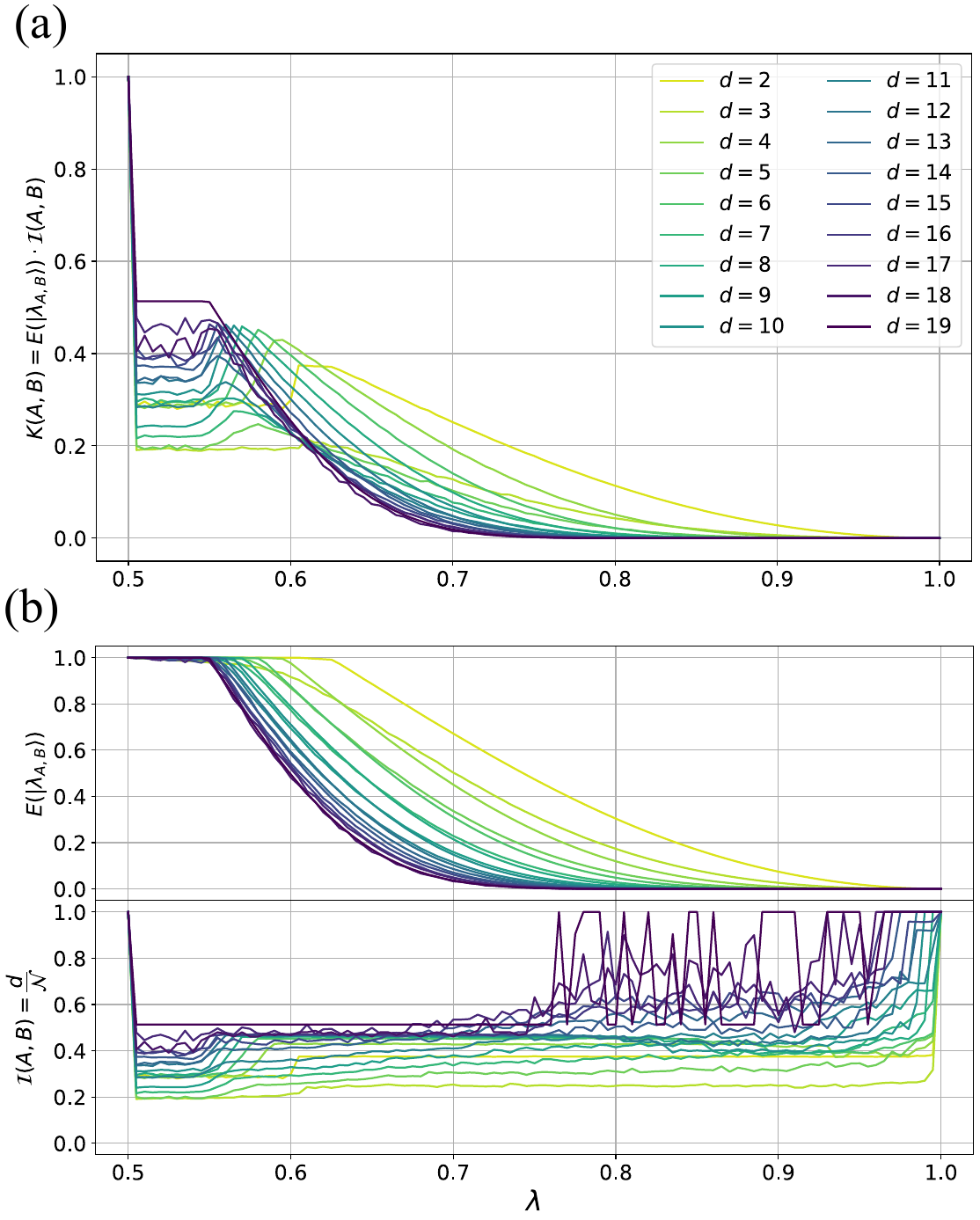}
  \end{subfigure}
  \hspace{5mm}
  \begin{subfigure}{0.45\textwidth}
      \label{fig:honeycomb_b}
      \centering
      \includegraphics[width=8.2cm]{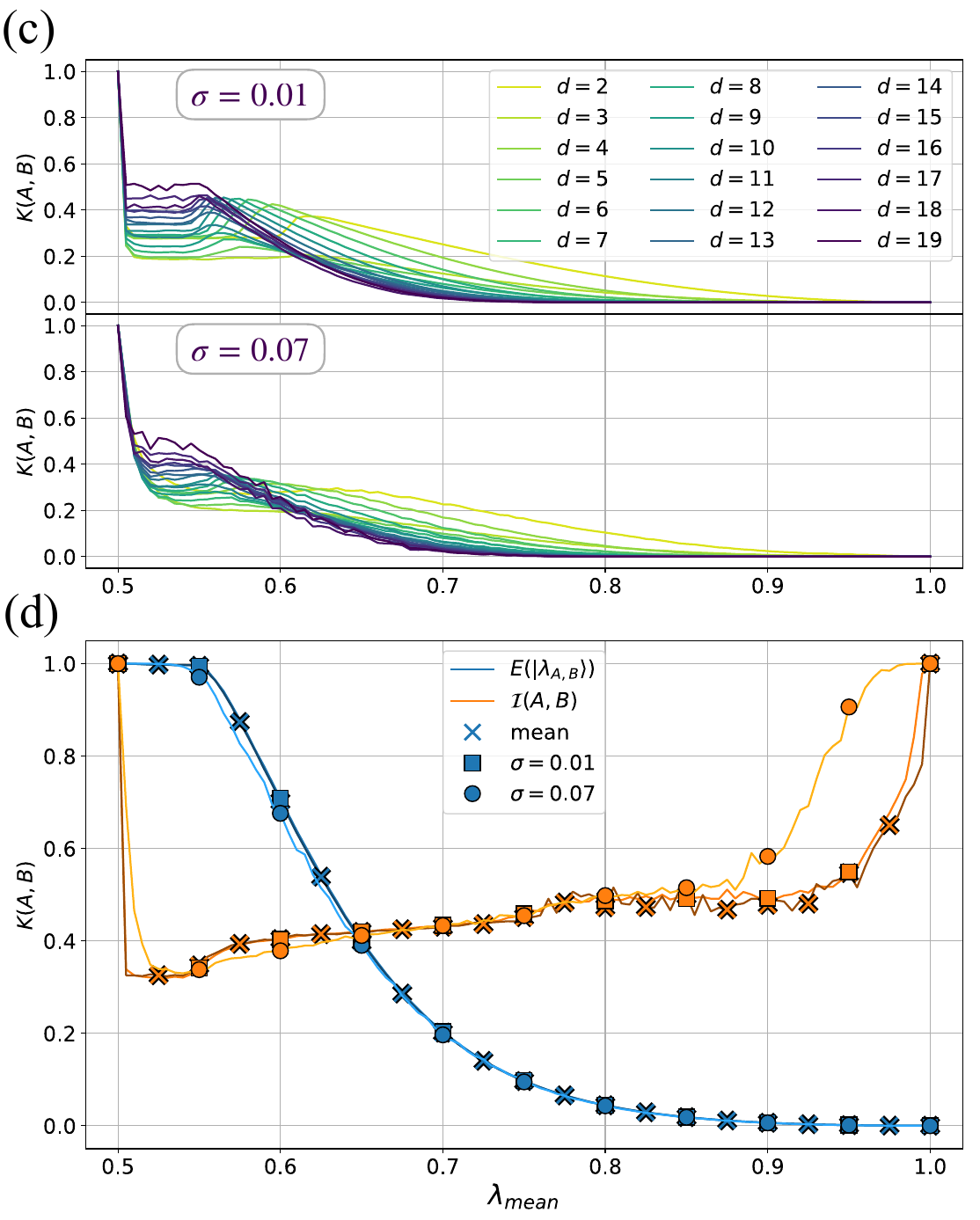}
  \end{subfigure}
  \caption{Results for a honeycomb lattice quantum network with 96 nodes arranged in 36 hexagons. \hyperref[fig:honeycomb_a]{(a)}~Connectivity of the network after quantum percolation between all possible distant node pairs, plotted against the Schmidt value assigned to all the initial states of the network. Each curve corresponds to the average connectivity of all pairs of nodes at a distance $d$. \hyperref[fig:honeycomb_a]{(b)}~Entanglement of the final state between distant nodes $A$ and $B$ (above) and integrity of the network (below) after quantum percolation between all possible pairs at distance $d$, plotted against the Schmidt value assigned to all the initial states of the network. \hyperref[fig:honeycomb_b]{(c)}~Connectivity of the network after quantum percolation between all possible pairs at distance $d$, plotted against the mean Schmidt value of the initial states of the network. The initial Schmidt values are drawn from a truncated normal distribution, with an enforcement on the mean $\lambda_{mean}$, with standard deviations $\sigma = 0.01$ (above) and $\sigma = 0.07$ (below). The results are averaged over 10 different network samples. \hyperref[fig:honeycomb_b]{(d)}~Comparison of entanglement and integrity between quantum networks with initial equal Schmidt values (mean), and drawn from a truncated normal distribution with mean enforcement, with standard deviations $\sigma = 0.01, 0.07$.}
    \label{fig:honeycomb_plots}
\end{figure*}

Fig.~\ref{fig:square_lattice_plots} presents the results for connectivity, entanglement, and integrity for a 10x10 square lattice. Compared to the diagonal square lattice, this topology contains significantly fewer states per unit cell, limiting the detectable local percolation strategy to the $2S,2D$ strategy. The Schmidt value threshold $\lambda_{th}^{2S,2D}$ associated with this strategy is indicated in the plot as a vertical red line. From Fig.~\hyperref[fig:square_lattice_plot_a]{10a}, we observe that the only curve that clearly obeys a transition at the percolation threshold corresponds to percolation for nodes at maximum distance. This is due to a limitation of the $2S,2D$ strategy in the square lattice, which can only produce maximally entangled states between nodes that lie along the same diagonal. In this topology, there exist only two pairs of nodes at maximum distance, both positioned along the same diagonal. Consequently, the connectivity shift is evident in the plot only for these maximum-distance pairs. The remaining curves represent an average of two types of node pairs: those aligned along the same diagonal, which exhibit the transition at the percolation threshold, and those positioned on different diagonals. For the latter, quantum percolation requires the distillation of multiple entanglement-swapping paths between the nodes. As illustrated in Fig.~\hyperref[fig:square_lattice_plot_a]{10b}, it is still possible to establish maximally entangled states between distant node pairs, provided the network has up to a certain initial average imperfection in its entanglement. However, the threshold at which entanglement decay begins varies depending on the distance between the nodes. This analysis highlights that while the simple square lattice serves as an instructive model, it is not ideal for large-scale experimental realizations due to its initial configuration constraints, which limit the generation of high-quality states between most distant node pairs. Finally, when analyzing networks with non-uniform initial entanglement (Fig.~\hyperref[fig:square_lattice_plot_b]{10d}), we notice that the entanglement of the final state remains mostly unaffected by the initial quenched disorder. However, the integrity of the network improves when the initial entanglement disorder is high, as indicated by a smoother transition compared to the sudden integrity shift observed in the uniform distribution case.

Fig.~\ref{fig:honeycomb_plots} illustrates the results for honeycomb lattice quantum networks with 36 hexagons. Similar to the square lattice case, the optimal percolation threshold varies depending on the distance between the node pairs. Moreover, the unit cells of this network topology involve non-local strategies between nodes, like the strategy associated with the Schmidt value threshold $\lambda_{th}^{2S+1SS,3D}$ defined in Sect.~\ref{sec:results}. Consequently, all the observed transition points marking a decay in the entanglement are associated to non-local strategies, whose thresholds are non-trivial to detect analytically. When comparing networks with uniform entanglement distribution to those with quenched disorder, we observe that the entanglement of the final state and the integrity of the network present similar behaviors in both cases, with the integrity transition remaining smoother in networks with higher initial disorder. 

\bibliography{bibliography}

\end{document}